# Multiconjugate Adaptive Optics for Astronomy

François Rigaut[1] and Benoit Neichel[2]

[1]Research School of Astronomy and Astrophysics, Australian National University, Canberra, ACT 2611, Australia; email: francois.rigaut@anu.edu.au

[2]Aix Marseille Université, CNRS, CNES, LAM, Marseille, France; email: benoit.neichel@lam.fr

**Keywords**

observational astronomy, astronomical instrumentation, tomography, atmospheric turbulence, point spread function

**Abstract**

Since the year 2000, adaptive optics (AO) has seen the emergence of a variety of new concepts addressing particular science needs; multiconjugate adaptive optics (MCAO) is one of them. By correcting the atmospheric turbulence in 3D using several wavefront sensors and a tomographic phase reconstruction approach, MCAO aims to provide uniform diffraction limited images in the near-infrared over fields of view larger than 1 arcmin$^2$, i.e., 10 to 20 times larger in area than classical single conjugated AO. In this review, we give a brief reminder of the AO principles and limitations, and then focus on aspects particular to MCAO, such as tomography and specific MCAO error sources. We present examples and results from past or current systems: MAD (Multiconjugate Adaptive Optics Demonstrator) and GeMS (Gemini MCAO System) for nighttime astronomy and the AO system, at Big Bear for solar astronomy. We examine MCAO performance (Strehl ratio up to 40% in H band and full width at half maximum down to 52 mas in the case of MCAO), with a particular focus on photometric and astrometric accuracy, and conclude with considerations on the future of MCAO in the Extremely Large Telescope and post–HST era.

## Contents

# 1. FROM NATURAL TO LASER GUIDE STARS

This paper is one in a series of *ARAA* adaptive optics (AO) reviews, starting with Beckers (1993) and more recently with Davies & Kasper (2012). In this review, we concentrate particularly on multiconjugate adaptive optics (MCAO). The reader is referred to the previous reviews for more details on general aspects of AO.

**Deformable mirror (DM):** small mirror whose shape can be adjusted to rectify changing optical aberrations such as those induced by atmospheric turbulence

## 1.1. History of AO

Since the early times of the telescope, astronomers realized that atmospheric turbulence degrades celestial images (there is a famous quote from Newton about this; see Tyson 2015, chap. 4), but Babcock (1953) was the first to see that something could be done to mitigate the problem. In his seminal AO paper, Babcock laid out essentially all the basic principles of the discipline: a wavefront sensor (WFS) to measure the wavefront aberrations induced by atmospheric turbulence; a deformable mirror (DM) to compensate for these aberrations; the closed-loop arrangement; and the limitations of the technique, including the fact that it was limited to the vicinity of stars bright enough to be used as guides, in effect a very small fraction of the sky. But in 1953, the technology was just not ready, so the idea remained a concept until the 1970s. At the heart of the cold war, the Defense Advanced Research Projects Agency (DARPA) started a program to build the first AO system in the hope, certainly, of being able to obtain better images of foreign satellites passing overhead. Although challenging, the developments were extremely successful and the Real-Time Atmospheric Compensator closed the first loop in the laboratory in 1973, followed by field tests and larger, faster, more capable systems (for a fascinating history of AO, see Hardy 1998).

**Laser guide star (LGS):** an artificial star created by use of a guide star laser

These developments advanced the state of the technology, and by the mid-1980s, AO components such as detectors, DMs, and (analog) reconstructors were available, albeit under restricted use. Having somehow heard about the DARPA AO program success, astronomers became interested in applying this technique to astronomy (for interesting insights, read McCray 2000). After a failed attempt at the National Optical Astronomical Observatory, the first astronomical system was developed in Europe under European Southern Observatory (ESO) coordination. COME-ON was a low-order system, with a 19-actuator DM and a 20-subaperture Shack–Hartmann WFS, and used a $32 \times 32$ pixel near-infrared (NIR) imager as the science detector. It saw first light at the Observatoire de Haute-Provence in 1989 and achieved the diffraction limit of the 1.5-m telescope in the NIR (Rousset et al. 1990).

After a major declassification of AO-related military research in the United States (Collins 1992, Duffner 2008), astronomical AO accelerated in the 1990s with development of larger systems (Beuzit et al. 1997) and alternative concepts like curvature sensing and compensation (Roddier & Roddier 1988, Roddier et al. 1991). See Roddier (1999) for a more detailed description of the first years of astronomical AO.

In the mid- to late 1990s, various teams worked on demonstrating AO with a laser guide star (LGS), first proposed by Foy & Labeyrie (1985). The Multiple Mirror Telescope in 1995 (Lloyd-Hart et al. 1995) and the 3-m Shane telescope at the Lick Observatory in 1996 (Max et al. 1997) were the first astronomical telescopes to demonstrate laser guide star adaptive optics (LGSAO) and were later on followed by Calar Alto (Davies et al. 2000). Initially, LGSAO brought only moderate image quality gains. It took many years and dedication to understand and overcome all the challenges associated with LGSs. The undisputed leader in that venture is the Keck Observatory,[1] which closed the loop in LGS mode in 2003, followed by the Gemini Observatory, the Very Large Telescope (VLT) and the Subaru Telescope, the Palomar Observatory, Robo-AO, Large Binocular Telescope (LBT), Southern Astrophysical Research Telescope, William Herschel Telescope, and others.

The same period saw the emergence and the demonstration of other key technologies like deformable secondary mirrors (Brusa-Zappellini et al. 1999) and pyramid WFSs (Ragazzoni 1996), as well as an explosion of new concepts, like ground-layer adaptive optics (GLAO), laser tomography adaptive optics (LTAO), MCAO, and others (see Section 1.5 and generally the proceedings of the conference "Beyond Conventional Adaptive Optics" held in Venice in 2001). MCAO is at the heart of this review. It was initially proposed by Beckers (1988, 1989) for its potential to increase the field of view (FoV) of AO systems. The idea involves using several DMs and WFSs to compensate for the turbulence in the volume of air above the telescope so that the compensation can be effected over a finite but extended FoV.

AO is now an important part of the astronomical instrumentation landscape. It is producing high-impact astronomical science (Davies & Kasper 2012) and has found its place between seeing-limited and space-based observatories. It has become particularly relevant in the era of the extremely large telescopes,[2] which are enabled by and conceived with AO systems fully integrated into the telescopes.

---

[1] According to Wizinowich (2013), the Keck LGS system has produced 72% of all LGS-based astronomical papers from 2004 to 2012.

### 1.2. Introduction to AO

The Sun dumps heat into the atmosphere. Through mechanical processes, in particular, wind, this heat creates turbulence, eddies of different temperatures, and thus different indices of refraction. A plane wave coming from a distant object will cross this turbulent media and get distorted, as it crosses "bubbles" of various indices of refraction. Part of the wavefront will get more or less delayed, and the wavefront will be corrugated when it hits the telescope. The image created from this aberrated wavefront will be distorted and, when averaged over many realizations, will lead to a blurry image, degrading the angular resolution. The size of this blur is called the "seeing." It is often characterized by the angular full width at half maximum (FWHM) of the seeing-limited images, which is given by $\lambda/r_0$, where $r_0$ is a characteristic length over which the wavefront is roughly flat ($\lambda/4$). Note that the seeing not only blurs the images but, by spreading the flux, also reduces the ability to detect faint objects against the sky background. At this point, it is important to remember that the theoretical angular resolution of an optical system is inversely proportional to the diameter of the aperture, following FWHM $= \lambda/D$, where $\lambda$ is the imaging wavelength and $D$ the aperture diameter.

The seeing at visible wavelengths is of the order of 1 arcsec in good sites and 0.5 arcsec in exceptional ones (Mauna Kea, Chilean sites, and a very limited number of other sites). In turn, the seeing will be the limiting effect at visible or NIR wavelengths on all large telescopes. Without this limitation imposed by the turbulence, these telescopes would be potentially capable of angular resolutions ten times better than what they can achieve from the ground.

One obvious solution to this problem is to place the telescope outside of the atmosphere, in space. Although this is the preferred solution, as it solves other problems like atmospheric filtering of some wavelength ranges, it is expensive and has more stringent technological limitations [e.g., the astronomy community is currently building 25–38-m telescopes for $1–2 billion—not including operation costs—while the 6.5-m *James Webb Space Telescope* budget approaches $8 billion].

Astronomers have been looking for a way to mitigate this seeing problem, and in 1953 Horace Babcock proposed the concept of AO. An example of this atmospheric turbulence compensation is presented in **Figure 1**, which shows a field in the globular cluster M13, without and with AO.

AO uses a combination of a WFS and a wavefront corrector to analyze and partially compensate for the wavefront aberrations in real time. There are many different types of WFSs, but their common function is to measure the distortion of the wavefront or the phase. Intensity is currently the only way to encode the wavefront as no detector is sensitive to phase in the optical/NIR domain. It remains a fundamental difference with detectors in the radio domain. The function of a wavefront corrector—we use the generic term DM in this review—is to correct for these distortions. It is generally a mirror the shape of which can be controlled by some kind of electromechanical effect; we call it an actuator. The WFS is generally located downstream of the DM, in a so-called closed-loop arrangement: When the WFS sees a deviation from the perfect wavefront, the DM shape is updated to cancel it. The DM commands are computed from the WFS error signal by a

---

[2]The ESO Extremely Large Telescope (ELT), the Thirty Meter Telescope, and the Giant Magellan Telescope are the three extremely large telescope projects currently being designed and built by international consortia.

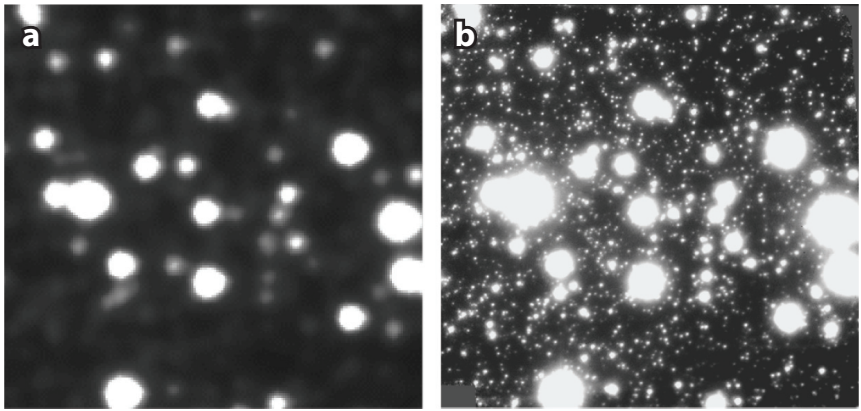

**Figure 1**

M13 observed at Gemini North (*a*) without and (*b*) with correction from the AO system Altair. The seeing-limited image in panel *a* was taken with the Acquisition camera at V-band, with a FWHM of 0.85 arcsec. The AO-corrected image in panel *b* is at K′ (2.12 μm) with a FWHM of 92 mas. The field of view is 22 arcsec. AO brings a gain in angular resolution (i.e., more details), as well as a gain in sensitivity: The light from stars is concentrated against a smaller patch of sky, increasing the signal-to-noise ratio for point sources. Figure adapted from Gemini Observatory/NSF/AURA/CNRC-Herzberg. Abbreviations: AO, adaptive optics; FWHM, full width at half maximum.

real-time computer (RTC), using a control matrix and assuming the system is linear. There are many methods to derive the control matrix (see Section 2.4 for an overview of the particular case of MCAO). Most are generally based on some kind of inversion of the interaction matrix—the calibration of how the WFS reacts to changes in the DM shape.

All WFSs sample the wavefront spatially in a finite number of points (typically one $r_0$ at the wavelength at which good correction is wanted), and they sample it at some frequency (typically a few hundred hertz). These two conditions, plus the requirement that the WFS needs a minimum number of photons, mean that there is a critical guide star brightness, called the limiting magnitude. These limitations are detailed in the following section.

## 1.3. Limitations of AO

The wavefront compensation effected by AO is never perfect—although it can get close in the case of extreme adaptive optics (ExAO). AO systems are coming with their own intrinsic limitations, mostly imposed by the nature of turbulence itself. These limitations are multiple (spatial, temporal, and angular) and are detailed below. A detailed description of turbulence and its effect in image formation has been given by Roddier (1981). For more details on AO-related errors, see Tyson (2015).

- **Spatial, a.k.a. the fitting error**: There is turbulence down to spatial scales of the order of millimeters. For apertures of a few meters, a perfect correction would mean hundreds of thousands of spatial degrees of freedom. Beyond the technological challenge, driving so many degrees of freedom means having as fine a measure of the input aberrations, which would imply splitting the light from the guide star into as many parts, driving the guide star brightness requirement up. In addition, the power spectrum of the input phase fluctuation goes down with increasing spatial frequency, which means that the benefit in correcting higher and higher spatial frequencies become less and less. In other words, there is an optimal spatial cutoff frequency, which depends on the seeing, the guide star brightness, and

**Real-time computer (RTC):** computes the DM actuator position updates based on a WFS error signal

**Guide star laser (GSL):** a sodium laser used to excite sodium atoms in the upper mesosphere to reemit light, creating a guide star more mundane considerations like the budget to buy DMs, and so forth. This is referred to as spatial error, as it is a spatial property of the system, and it is often referred to as fitting error. The associated phase error, for a cartesian actuator geometry, is $\sigma_f^2 = 0.23 \, (d/r_0)^{5/3}$,[3] where $r_0$ is the Fried parameter, and $d$ is the DM effective actuator pitch.

- **Temporal, a.k.a. the servo lag error**: In an AO system compensation loop, there is necessarily a delay between the measurement by the WFS and the application of the correction by the DM. The minimum delay is the WFS exposure time. To this, more often than not, additional delays due to the WFS read out, the command calculation, etc., will add up. This means that by the time the correction is applied to the DM, the input perturbation will have changed, resulting in an imperfect correction. This is called servo lag error. The characteristic evolution time of the atmosphere is $\tau_0$, and the contribution of this error to the phase variance is $\sigma_t^2 = K(t/\tau_0)^{5/3}$, where $t$ is the loop sampling time and $K$ depends on the characteristics of the loop and control law.
- **Angular, a.k.a. the anisoplanatic error**: The turbulence is a 3D medium. In a given direction on the sky the measured phase is the 2D projection of the turbulent volume along the line of sight. Away from this line of sight, the phase changes because the beam does not cross the same volume of turbulence. Using one to correct the other results in an error $\sigma_a^2 = (\theta/\theta_0)^{5/3}$. This is called anisoplanatic error.

The combination of the limiting magnitude and the anisoplanatism angle $\theta_0$ leads to the concept of sky coverage, which is the fraction of the sky over which a suitable AO correction can be achieved. Typical numbers (limiting magnitude $R = 15$ and $\theta_0 \approx 30$ arcsec at 2 μm) lead to sky coverage of approximately 1%, a very low number indeed. This impaired the use of AO and led to the concept of artificial guide stars.

### 1.4. Laser Guide Stars

To circumvent the limitations imposed by the availability of a natural guide star (NGS), several solutions were proposed to create artificial guide stars. Linnick (1957) proposed to fly planes with lights where a guide star was needed. A more tractable method—although by no means trivial—was proposed by Foy & Labeyrie (1985). The method makes use of lasers, termed a guide star laser (GSL), to excite sodium atoms located in the upper mesosphere. These atoms will then reemit light at the same wavelength, forming an LGS. The sodium layer is replenished by meteorites. The conditions of temperature and pressure are such that the layer is relatively well defined: Below 90 km, the sodium atoms combine with other sodium atoms or other species to form molecules, forming a well-defined boundary. The upper limit of the sodium layer is defined by the recombination of ablated sodium ions and free electrons from the ionosphere and by the cross section for meteor ablation (Pfrommer & Hickson 2010). The combination of the density of sodium atoms and the very large cross section of the sodium D2 line at 589 nm (the one used in orange streetlights) makes this the most favorable choice.

An alternative method, also relying on lasers, is to use the Rayleigh backscatter from air molecules ($N_2$, $O_2$, etc.). Because the backscattering medium is continuous, a combination of pulsed lasers and time-gated detectors has to be used. Due to the exponential decay of atmospheric density with altitude, the maximum height at which a Rayleigh guide star can be created is approximately 20 km (Fugate 1991, Hart et al. 2010).

---

[3]The description of turbulence by Kolmogorov leads to this unusual exponent; many atmospheric turbulence parameters vary as $\lambda^{6/5}$: $r_0$, $\tau_0$, $\theta_0$, $d_0$.

> **STREHL RATIO**
>
> The Strehl ratio $\mathcal{S}$ is a measure of how close to the theoretical diffraction limit the image is. The Strehl is the ratio of the maximum intensity of an actual image to the maximum intensity of a fully diffraction-limited image normalized to the same total flux—an Airy pattern has a Strehl ratio of 100%, whereas the typical Strehl ratio of seeing-limited images in the NIR and on an 8-m telescope is a few percent. Strehl is a good indicator of AO performance for $\mathcal{S} > 15\%$, below which the FWHM becomes more relevant, as the Strehl ratio plateaus.

Many telescopes are now equipped with LGSs, whether they are Rayleigh or sodium; at Mauna Kea, one can often spot four orange beams crisscrossing the sky (Goebel 2017). LGSs are also an integral component of all the extremely large telescopes being currently designed. Sodium guide stars do not come without limitations themselves, which are presented below.

**1.4.1. Cone effect.** The light from an LGS comes from a point at a finite distance, and thus forms a cone that does not overlap exactly with the cylinder of light coming from an astronomical object, which is located, for all practical purposes, at infinity. The rays coming from the LGS are thus not crossing the turbulent volume at the same location as the rays from an NGS. This is called the cone effect (Tallon & Foy 1990) or sometimes focal anisoplanatism (Fried & Belsher 1994). It becomes more severe for shorter-range LGSs or for larger telescope apertures. Fried & Belsher (1994) derived the amplitude of the phase error associated with cone effect as $\sigma_\varphi^2 = (D/d_0)^{5/3}$, where $d_0$ is a parameter that can be computed from the $C_n^2$ profile and the altitude of the LGS, and can be viewed as the aperture diameter for which the cone effects start to be significant [1 rad$^2$ of phase error leads to a Strehl ratio (see the sidebar titled Strehl Ratio) of approximately 37%]. The length $d_0(0.5\ \mu m)$ is typically a few meters for an LGS at 90 km or a fraction of a meter for a Rayleigh LGS and varies as $\lambda^{6/5}$.

**1.4.2. Tip-tilt indetermination.** Rigaut & Gendron (1992) describe the tip-tilt (TT) indetermination problem for LGS single conjugate adaptive optics (SCAO) systems: The position of the LGS is polluted by the uplink wander of the laser beam, which makes the LGS position useless to derive the downlink TT.[4] This problem can be mitigated by the use of an NGS to sense TT. The NGS can be significantly fainter than it can for regular high-order NGS systems, as the entire aperture can be used to collect the TT information, leading to a magnitude gain. For a Shack–Hartmann WFS, this gain can be approximated by $2.5 \times \log 10(N_{sub})$, where $N_{sub}$ is the number of subapertures.

**1.4.3. Low-order aberrations.** The atomic density distribution within the sodium layer changes constantly, consequently affecting the sodium return flux and its vertical distribution (Pfrommer & Hickson 2010). As seen from the telescope focal plane, the LGS is a 3D extended object (similar to a cigar), with approximately a Gaussian distribution of FWHM $\approx 1.5$ arcsec in the $(x, y)$ plane parallel to the telescope pupil and a complex distribution versus altitude. Variations of the sodium layer altitude directly translate into focus errors for LGSAO systems. The resulting wavefront error is proportional to the altitude variation and scales as $D^2$. For instance, for a 30-m telescope observing at zenith, a change of 1 m in the sodium layer altitude results in a 4-nm wavefront error. Sodium altitude fluctuations cannot be distinguished from atmospheric focus changes, and an NGS is required to disentangle these two effects. For 8-m-class telescopes, the focus sensing on an NGS to compensate for sodium layer altitude drift has to be done typically on a minute timescale, hence not impacting the sky coverage (Neichel et al. 2013). Because of the $D^2$ dependency, for extremely large telescope–sized telescopes, the focus control has to be done at a few tens of hertz, and dedicated low-order NGS systems must be developed. As an additional effect, the finite height and structure variations of the sodium layer can create quasi-static aberrations only seen by the LGS path (Clare et al. 2007), creating a new source of noncommon path optical aberrations that have to be measured by a dedicated low-order WFS (the so-called reference or truth WFS).

---

[4]Note that the uplink and the downlink are not the same, as—for various reasons—the full aperture is never used to project the uplink laser beam. If this was done, the LGS would actually appear totally motionless, as the uplink and downlink beam deviations would cancel each other out.

### 1.5. Taxonomy of AO Species

Physical limitations imposed by anisoplanatism and the signal-to-noise ratio in the WFS make it impossible to obtain a perfect image over an arbitrarily large FoV. Choices have to be made. This has spawned an entire zoo of AO concepts, which span different subspaces in the Strehl ratio/sky coverage/FoV volume: SCAO, LGSAO, GLAO, MCAO, LTAO, ExAO, and multiobject adaptive optics (MOAO) (see **Figure 2**).

SCAO is the classical AO, which uses a single guide star and single DM. This is where it all began in the 1970s in the United States (DARPA programs), and then the early 1990s in Europe (COME-ON, i.e., CGE Observatoire de Meudon ESO ONERA). Mature examples of these include NAOS (Nasmyth Adaptive Optics System; Rousset et al. 2003), Altair (ALTtitude Conjugate Adaptive Optics for the Infrared; Herriot et al. 2000), the Keck NGS system (Wizinowich et al. 2000), the Subaru AO188 (Minowa et al. 2010), and the LBT system (Esposito et al. 2010, 2011).

LGSAO adds an LGS (sodium or Rayleigh) to SCAO, drastically improving the sky coverage (up to approximately 30%, from 1% for NGS SCAO), at the price of leading to more complex systems (Wizinowich et al. 2006). In addition, the performance of LGSAO systems is impacted by the cone effect and still needs an NGS for TT compensation.

ExAO is essentially SCAO on steroids, providing an extremely high order of correction on very bright guide stars over a small FoV to give access to very high contrasts required for exoplanet or debris disk imaging. The challenge in ExAO is that control of the very tight error budget (typically a total RMS error of 60–80 nm in current systems) requires every term to be controlled with the utmost attention. ExAO requires many actuators, and only works with NGS, as the cone effect disqualifies LGSAO. Due to the strong motivation in imaging the first exoplanets, there are a number of ExAO systems currently pushing the contrast envelope: SPHERE (Spectro-Polarimetric High-Contrast Exoplanet Research Instrument; Beuzit et al. 2008, Fusco et al. 2014), GPI (Gemini Planet Image; Macintosh et al. 2014), SCExAO (Subaru Coronagraphic Extreme Adaptive Optics; Martinache & Guyon 2009, Martinache et al. 2014), PALM-3000 (Palomar Adaptive Optics System; Dekany et al. 2013), and MagAO (Magellan Adaptive Optics; Close et al. 2013).

At the other extreme, GLAO provides modest FWHM gains over FoVs of up to 15 arcmin (Rigaut 2001, Tokovinin 2004). It does so by compensating for the ground layer of turbulence, where the majority of turbulence is located at most sites. The compensation is often done by an adaptive secondary mirror which is approximately conjugated to the ground. Because it is done in the telescope aperture and conjugated to where the turbulence occurs, the correction is valid over very wide FoVs. Tokovinin (2004) derived the theoretical performance of GLAO. Baranec et al. (2009) and Rabien et al. (2017) report on results of systems using Rayleigh guide stars at

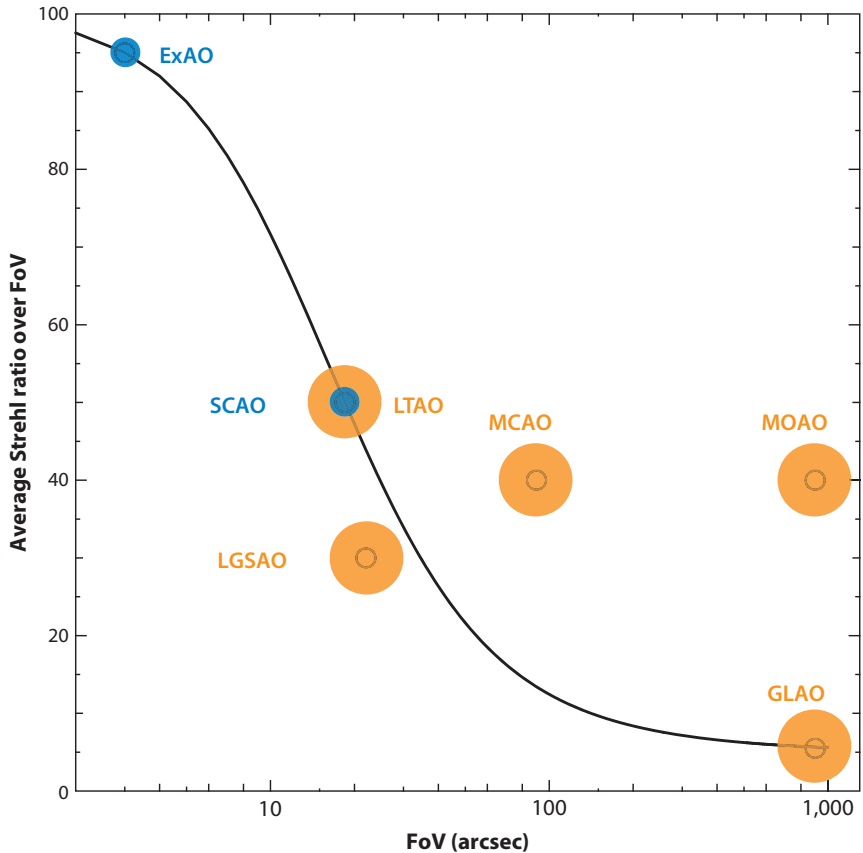

**Figure 2**

Trends in coverage of the (Strehl, FoV) space by the various adaptive optics species for the H band. Blue points are NGS-based concepts only, with sky coverage from 0.001% to ≈1%. Orange points are LGS-based concepts, with sky coverage of 30–70%. The line traces the performance limit imposed by anisoplanatism in the case of a single DM. LGSAO is below the line because of the cone effect. MCAO is above the line because it uses several DMs, beating the anisoplanatism limitation—but it is limited by generalized fitting; thus it does not reach the SCAO level of performance in on-axis systems. MOAO is plotted as accessing a very large FoV, but it has to be realized that the MOAO buttons only cover a limited number of very small fields. This diagram is illustrative only and roughly corresponds to currently doable systems for near-IR wavelengths; actual detailed performance would depend on an extensive series of assumptions. Abbreviations: DM, deformable mirror; ExAO, extreme adaptive optics; FoV, field of view; GLAO, ground-layer adaptive optics; LGS, laser guide star; LGSAO, laser guide star adaptive optics; LTAO, laser tomography adaptive optics; MCAO, multiconjugate adaptive optics; MOAO, multiobject adaptive optics; NGS, natural guide star; SCAO, single conjugate adaptive optics.

the Multiple Mirror Telescope and at the LBT. GLAO is often described as a seeing improver, transforming poor- into median-condition nights, median into good, and good into exceptional nights. As a rule of thumb, a gain of a factor of two is obtained in FWHM and encircled energy in the NIR. This is almost universally predicted by simulations (Le Louarn et al. 2006, Hart et al. 2010), and to date seems to be confirmed to include the ESO adaptive optics facility (AOF) (Arsenault et al. 2012) and Subaru Ultimate (Minowa et al. 2017). Most extremely large telescope projects have a GLAO mode in the plan.

One major limitation of LGSAO is the cone effect (see Section 1.4.1). LTAO is a concept that mitigates, or entirely solves, the cone effect limitation. It does so by using several LGSs to probe the cylindrical volume of turbulence in the telescope line of sight. The phase in the direction of the object of interest is obtained through a tomographic reconstruction process (Lloyd-Hart et al. 2006). This can lead to a high Strehl ratio, but because only one DM is used, it suffers from the same anisoplanatic limitations as SCAO.

MOAO (Hammer et al. 2004) is somewhat similar to LTAO in the sense that many guide stars (NGS or LGS) are used, through a tomographic process, to derive the command to apply to a DM to correct in one particular direction. The difference is that the many guide stars cover a much larger FoV—many arcminutes—and that the DM is downstream from the sensing stage; ideally integrated into a small unit staring at the astronomical object of interest and additionally including some kind of science imager or spectrograph. The intent is to have many of these DM/science instruments units to bring a multiplex advantage. One could imagine, for instance, having integral field units pointed at galaxies with each integral field unit also having its own integrated DM. MOAO provides small patches of good correction scattered across a large field and is well suited to, e.g., extragalactic studies. The main challenge of MOAO is related to the open-loop operation, which imposes new linearity, dynamical range, and hysteresis requirements on the WFSs and the DMs. The cost associated with multiplexing—that is, multiplying the number of pick-offs, DMs, and spectrographs—is also an issue. MOAO has been demonstrated recently at the William Herschel Telescope (Gendron et al. 2011) and Subaru (Lardière et al. 2014).

**Figure 2** summarizes the location of all AO concepts in a Strehl versus FoV representation. Davies & Kasper (2012) spend a considerable amount of effort describing these various AO concepts. The reader is referred to their review for more details.

## 2. OPENING THE FIELD WITH MCAO

### 2.1. Principles

The principles of MCAO are presented in **Figure 3**. Several DMs are stacked in a series to form a 3D corrector, which can then be optically conjugated to the whole turbulent volume and thus provide anisoplanatic correction (Dicke 1975, Beckers 1988, Ellerbroek 1994).

A piece of optics A is said to be optically conjugated to a layer X when X is imaged (or in focus) on A. Imagine a mirror in the AO instrument, optically conjugated to 10 km above the telescope: If a plane were to pass in the beam at 10 km above the telescope, the image of the plane would be perfectly in focus on this mirror. What works for amplitude also applies to phase, so that a DM optically conjugated to some layer will be able to exactly compensate for phase aberrations occurring in this layer.

The information on the 3D volume of turbulence is provided by multiple guide stars (NGS or LGS) and a tomographic processor. Tomography can be simply described as a mathematical method to reconstruct 3D content based on multiple projections of the 3D content onto 2D measurements from different angles. In MCAO, the information sought is the phase in the turbulent 3D volume above the telescope based on 2D phase measurements made in a number of single directions. The mathematical and algorithmic techniques associated with tomography are described in Section 2.4.

MCAO was demonstrated at ESO with NGSs (Marchetti et al. 2008), and more recently, at Gemini with LGSs (Rigaut et al. 2014, Neichel et al. 2014c), providing moderate Strehl (30 to 50%) over an FoV of 2 square arcmin, which is an order of magnitude larger than with classical AO. As a typical example of the MCAO performance, **Figure 4** shows NGC 288 in the H band (1.65 μm) over an FoV of $87 \times 87$ arcsec$^2$. The image, obtained during Gemini MCAO System (GeMS) first light, displays an FWHM of approximately 80 mas over the entire image, with an rms of 2 mas.

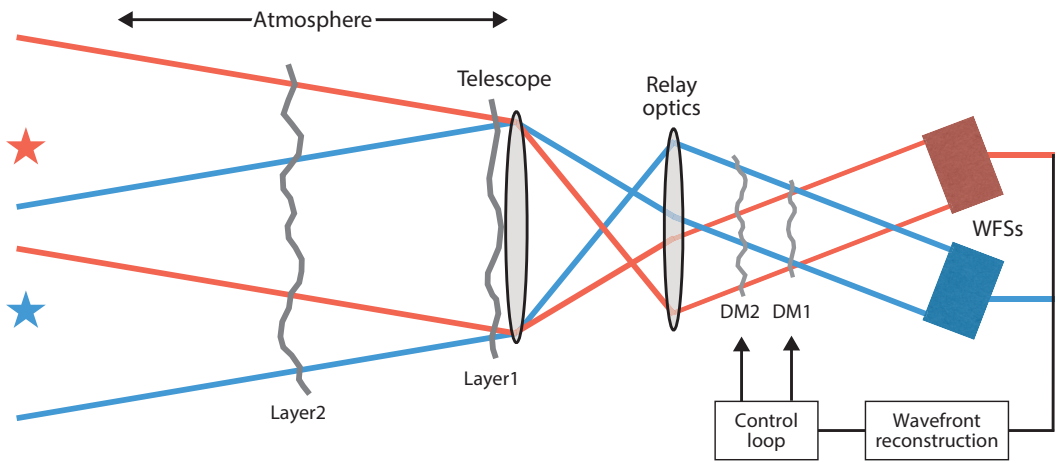

**Figure 3**

Principles of MCAO. By probing the atmospheric turbulence in various directions using multiple WFSs, and using tomographic techniques, MCAO can reconstruct the turbulence in 3D (atmospheric turbulence is mostly confined between 0 and 15 km). The phase correction is then projected onto a limited number of DMs, optically conjugated/located at various altitudes above the telescope. Theoretically, one could enlarge the field of view to arbitrarily large values. However, the presence of turbulence in between the DM locations limits the field of view practically acceptable. Adapted from Ellerbroek (1994). Abbreviations: DM, deformable mirror; MCAO, multiconjugate adaptive optics; WFS, wavefront sensor.

As a reference, the blue box gives an example of a typical FoV for an SCAO system (e.g., Altair, $20 \times 20$ arcsec$^2$), beyond which the anisoplanatism is starting to affect performance.

### 2.2. Scientific Motivation

There are several motivations to develop MCAO systems, and applications can be found in all astronomical areas from observations of the Sun to solar system objects (see **Figure 5**) to star clusters and extragalactic science. From a general point of view, there are potentially three main reasons driving MCAO observations: the need for the FoV, the need for sky coverage, and the need for astrometric accuracy.

The first category (the need for FoV) addresses spatially resolved and extended targets, like solar systems planets, globular clusters, star-forming regions, planetary nebulae, or some local galaxies. It may also be used as a multiplex advantage for extragalactic studies, for instance for galaxy clusters or strong lensing.

The second category of observations (the need for sky coverage) would cover small objects, i.e., objects not filling the wide corrected FoV but that would not be accessible at high angular resolution otherwise. Basically, all the targets that would be lying at an angular distance larger than 30 or 40 arcsec from a bright (e.g., $R < 16$) NGS would suffer from tilt anisoplanatism in classical LGSAO. The expected performance will then be drastically limited by the residual TT anisoplanatism: typical isokinetic angles (i.e., the distance from the TT guide star over which the Strehl is reduced to 37% of its on-axis value) are around 40 to 60 arcsec. Thanks to the multiple NGSs, an MCAO system can overcome tilt anisoplanatism and deliver uniform performance over a larger fraction of the sky than classical LGSAO. Targets that would fall into that category include QSOs, AGNs, isolated galaxies, neutron stars, pulsars, etc.

Finally, astrometry is quickly becoming one of the most in-demand science cases for MCAO, with a growing fraction of programs asking for good astrometric performance (statistics from NGC 288 with GeMS in the H band. One of the first light images from GeMS, this illustrates the gain in FoV brought by MCAO: The field is 87 × 87 arcsec$^2$ in area, almost 20 times the typical FoV of an SCAO system (e.g., Altair, 20 × 20 arcsec$^2$, in *blue*). The SCAO inset (*middle panel* on *right*) was generated assuming that the bright star in the upper right corner was used as a guide star. Figure adapted from Gemini Observatory/NSF/AURA. Abbreviations: AO, adaptive optics; FoV, field of view; GeMS, Gemini MCAO System; MCAO, multiconjugate adaptive optics; SCAO, single conjugate adaptive optics.

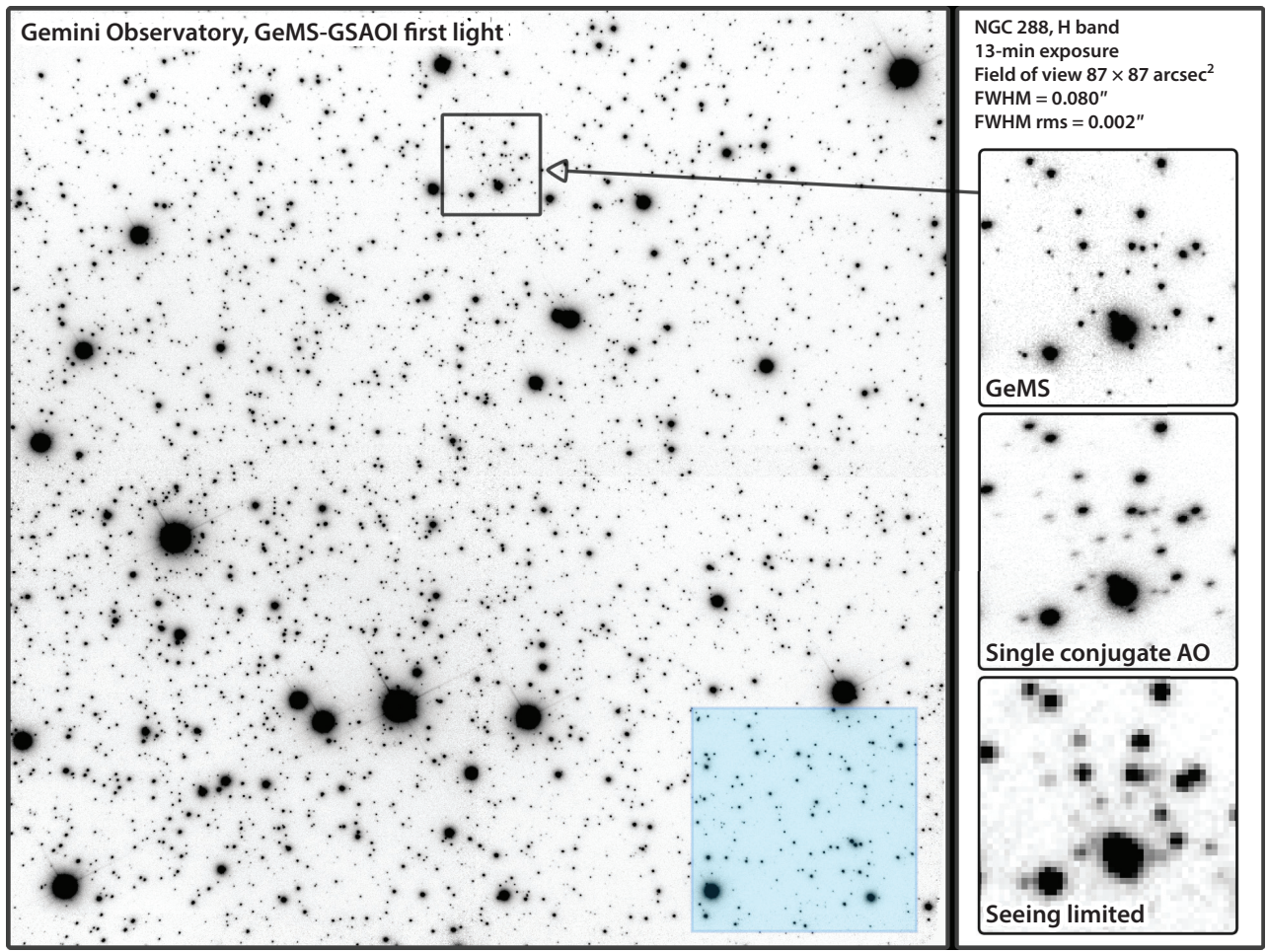

**Figure 4**

GeMS, as it is the only MCAO system in regular use for science operations). It is also one of the main science drivers for building the next generation MCAO systems for future extremely large telescopes. There are several reasons why astrometry is appealing with MCAO. First of all, and thanks to the deformable mirrors conjugated in altitude, an MCAO system is able to compensate for the field distortions induced by atmospheric turbulence. Indeed, optical aberrations conjugated outside of a pupil plane introduce field distortion, which will be changing all the time, as the turbulence does. These atmospheric-induced distortions are zero-mean; however, the convergence time required to reduce them below a given threshold may be long. An MCAO system, dynamically compensating for part of the altitude turbulence will naturally reduce the amount of distortion.

The second main reason why MCAO is attractive for astrometry is that it provides a large number of reference sources, and those sources have good image quality. As a first proxy, the centroiding error scales as the FWHM of the star. The gain brought by MCAO is then obvious: The wide FoV allows for a large number of uniformly high-quality reference stars. As seen in

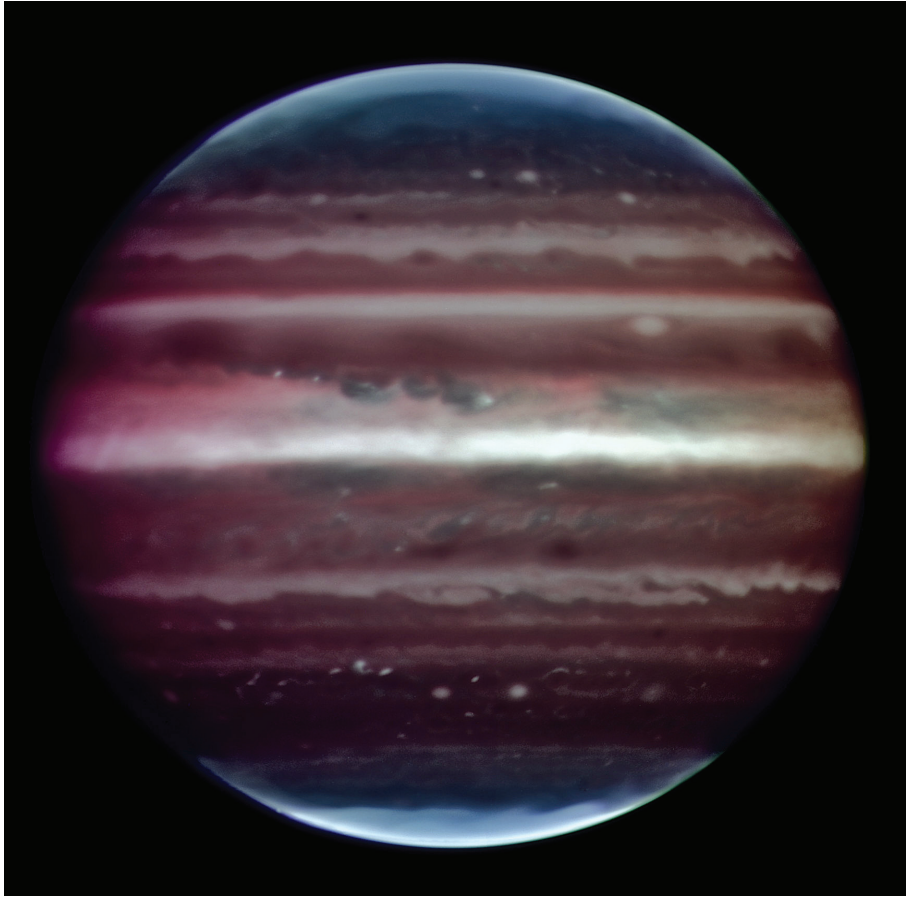

**Figure 5**

Jupiter as seen by MAD, an MCAO demonstrator built by the ESO in 2007. The planet has been imaged through three different filters (2, 2.14, and 2.16 μm), with an angular resolution of about 90 mas (corresponding to about 300 km resolution). By a comparison of this image with previous HST near-IR images, an alteration in the brightness of the equatorial haze was evidenced. The MAD images showed an increased sunlight-reflected haze, which could be interpreted either as an increase of the haze amount or as a motion of haze toward higher altitudes. See **https://www.eso.org/public/news/eso0833/** for more details. Figure adapted from ESO/F. Marchis, M. Wong, E. Marchetti, P. Amico, S. Tordo. Abbreviations: ESO, European Southern Observatory; MCAO, multiconjugate adaptive optics; MAD, Multiconjugate Adaptive Optics Demonstrator.

**Figure 4**, the area corrected by an MCAO system is typically 10 to 20 times larger than in classical LGSAO systems, potentially providing a much larger number of astrometric reference sources. A notable example is the Galactic Center. Current state-of-the-art observations are using masers located nearby the Galactic Center as absolute astrometric reference points (Yelda et al. 2010). However, those Masers are located within 40 to 60 arcsec from SgrA*, which makes mosaicing mandatory. With its large FoV, an MCAO instrument directly covers the whole field and removes the need for extra calibrations.

Another scientific motivation to use MCAO instruments is the uniformity of the corrected point spread function (PSF). A uniform PSF vastly improves the accuracy of the image analysis. It is the universal experience of AO users that data reduction is a critical problem, because of (*a*) the lack of proper and simultaneous PSF calibration and (*b*) PSF spatial variability in the field. Having a large,

**Point spread function (PSF):** characterizes the distribution of intensity in the focal plane of an image of a point source through an optical system uniform field goes a long way toward solving this problem: If a star is present in the FoV, it can be used for the whole MCAO-corrected field. Even for high galactic latitude fields usually lacking bright stars, the probability of having at least one H < 19 star in a 1 arcmin² field is high (60%).

In order to get a better overview of the science produced by MCAO systems, one can look at the papers that have been published based on Multiconjugate Adaptive Optics Demonstrator (MAD) and GeMS—the only two MCAO systems for nighttime astronomy that have produced science so far. However, this analysis is biased, as current MCAO systems on 8- to 10-m telescopes are to date only equipped with NIR imagers, to the exclusion of spectrographs, planned for a future phase. Looking at the literature for the science produced with 8-m MCAO systems, the following themes emerge:

- star clusters,
- astrometry,
- some extragalactic, and
- a bit of everything else.

The first two items are by far the main topics of the MCAO science, with around 70% of the papers published so far.

The star cluster science can actually be subdivided into two main areas: globular clusters (GCs) on one hand and young clusters or star-forming regions on the other hand. GCs are perfect candidates for MCAO observations as they usually cover a field of around 1 arcmin, and the high density of stars is a perfect application to illustrate the gain brought by AO corrections (Fiorentino et al. 2016). As an illustration, **Figure 6** shows a J, $K_s$ image of NGC 6624 as observed with GeMS (Saracino et al. 2016). The sensitivity gain for point sources observed with AO when compared with seeing-limited observations has been well illustrated in **Figure 1** and can reach up to 3 mag for crowded regions. The detection level is significantly improved, both because of the point source sensitivity gain and because of crowding reduction. As such, there has been a large number of published papers focusing on GC science by, among others, Momany et al. (2008), Ferraro et al. (2009), Moretti et al. (2009), Ortolani et al. (2011), Fiorentino et al. (2012), Turri et al. (2015), Saracino et al. (2015, 2016), Massari et al. (2016a), and Santos et al. (2016). One of the most impressive results is probably the double stellar population with different iron contents and ages highlighted by Ferraro et al. (2009) within the Terzan 5 system. It is also interesting to highlight that MCAO observations brought original observational ways to constrain the GC absolute ages (Bono et al. 2010), based on the difference in magnitude between the main sequence turnoff (MSTO) and a well-defined knee located along the lower main sequence. The lower main sequence–MSTO distance is a promising indicator of age because, unlike the distance between the MSTO and the horizontal branch or the red giant branch, it is only marginally dependent on the metallicity of the cluster, resulting in a higher age accuracy.

In the context of star cluster science, astrometry plays a special role and is usually identified as a specific goal for the observations. To a first level, adding proper motion information to the photometry can be used to clean the color-magnitude diagram from foreground and background stars (Lu et al. 2014, Fiorentino et al. 2016, Massari et al. 2016b). Astrometric accuracy to the level of a few milliarcseconds is already very helpful in this task. If the astrometric accuracy can be improved to submilliarcsecond levels, then proper motions to trace the GCs in space and time (using cluster age) become possible. Finally, if the astrometry accuracy is even better, this opens the door to internal dispersion: occurrence of intermediate-mass black holes at the center of Galactic GCs (Kains et al. 2016).

The star cluster science also addresses star-forming regions, and MCAO observations have been very productive in this area. One particular field of interest obviously is Orion, observed by

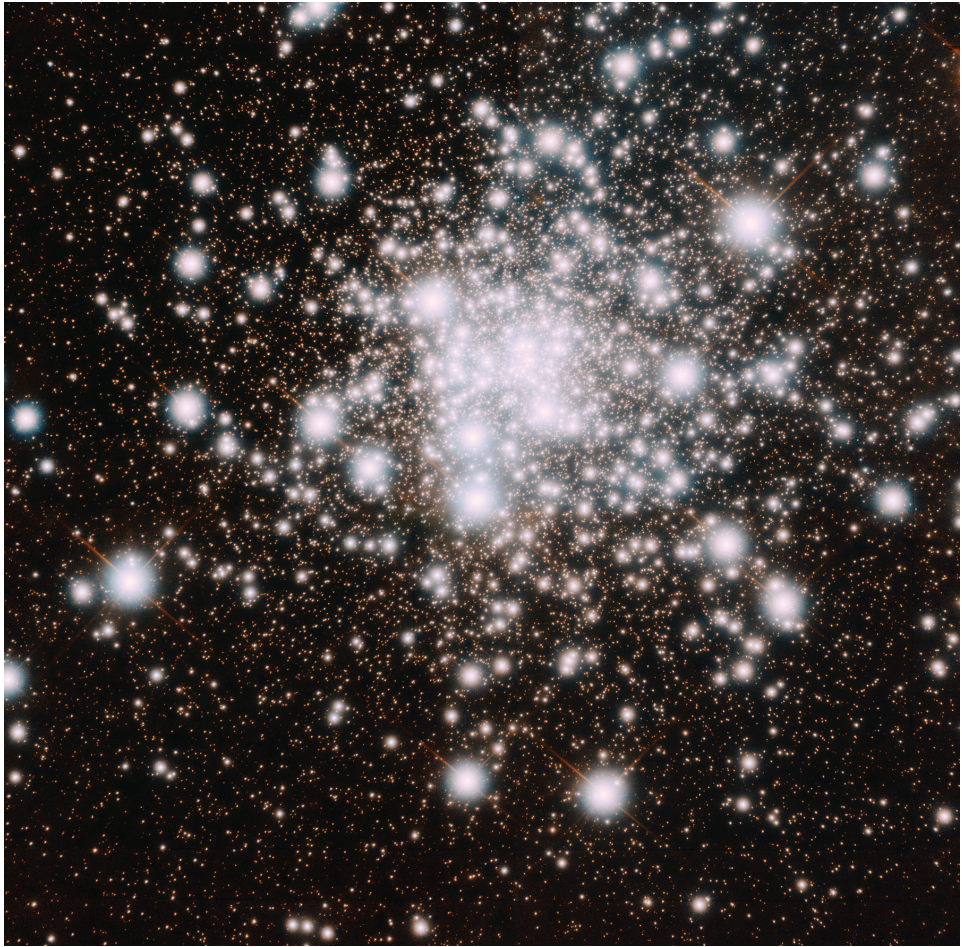

**Figure 6**

The globular cluster NGC 6624 with GeMS (Saracino et al. 2016). The field of view is 93 × 93 arcsec$^2$. This image combines two 420-s exposures in J and $K_s$ bands. AO particularly shines on globular clusters, and MCAO provides the additional advantage of a larger field of view and a more uniform PSF, resulting in more complete and stable photometry. Turri et al. (2015) report on combining HST data with GeMS data to produce the deepest ground-based photometry in a crowded field. The halo around the stars is a feature of AO correction—a consequence of the wavefront high spatial frequencies not being corrected. Figure adapted from Gemini Observatory/AURA/NSF. Abbreviations: AO, adaptive optics; GeMS, Gemini MCAO System; MCAO, multiconjugate adaptive optics; PSF, point spread function.

MAD and GeMS (Bouy et al. 2008, 2009a,b; Bally et al. 2015; Eisner et al. 2016). Combining the GeMS images with previous AO images, Bally et al. (2015) have been able to derive a 3D dynamical model of the region. Other examples of young clusters include Wd1 (Andersen et al. 2015), 30 Doradus (Campbell et al. 2010, Crowther et al. 2010), Trumpler 14 (Rochau et al. 2011), RCW 41 (Neichel et al. 2015), or even extragalactic regions in the Large Magellanic Cloud (LMC; Fiorentino et al. 2011, Bernard et al. 2016) and outside the Local Group (Gullieuszik et al. 2008).

There have been recently a couple of papers making use of the astrometry performance only. In particular, we cite Garcia et al. (2017), who derived the individual component masses of Luhman 16AB, a brown dwarf binary located at 2 pc from the Sun, or on a completely different scientific

**Asterisms:** compact sets of stars forming a small pattern on the sky topic, Fritz et al. (2017), who derived the absolute proper motion of Pyxis, a faint, distant halo cluster, aiming at bringing new constraints on the Milky Way mass. It is worth noting that NIR MCAO astrometry studies are fully relevant even when entering the *Gaia* era (Gaia Collab. et al. 2016), as they provide complementary information for very embedded regions (like the Galactic Center) and/or very crowded regions (like GCs) in which *Gaia* is blind.

As said above, only a handful of papers have been produced for extragalactic science (Liuzzo et al. 2013; Neichel et al. 2014a; Schirmer et al. 2015, 2017; Gibson et al. 2016; Sweet et al. 2017). One of the main reasons is likely because, NIR ground observations being severely limited by sky background, long exposures are required to reach faint targets. For instance, Schirmer et al. (2015) reported a 5σ detection for extended sources of $K_s = 25.6$ lim AB mag, for a total of about 15,000 s exposure on the Gemini 8-m telescope. Compared with seeing-limited observations, the signal-to-noise ratio gain provided by MCAO for extended objects is on the order of 2 to 3 (Schirmer et al. 2017). Furthermore, and before the *James Webb Space Telescope* (Gardner et al. 2006) becomes operational, MCAO-assisted imagers are the only ground-based instruments that can provide wide-field high angular resolution observations at K band.

MCAO observations can also benefit isolated targets, taking advantage of the improved sky coverage. Among others, we cite the work done on Y dwarfs (Leggett et al. 2015, Opitz et al. 2016), supernovae in nearby ultraluminous infrared galaxies (Ryder et al. 2014), pulsars (Zyuzin et al. 2013), planetary nebulae (Manchado et al. 2015), the protostar (Reiter et al. 2015), or isolated neutron stars (Mignani et al. 2008). Finally, for solar observations, MCAO is also strongly motivated by fast processes over large regions of the Sun, such as a big flare. Solar timescales are of the order of 10 s.

### 2.3. Sky Coverage and Natural and Laser Guide Stars

Does MCAO necessarily need LGS or can it be done with NGS? This point has seen harsh debates (Ragazzoni 2000): MAD, the first ever MCAO system on-sky, was using NGSs (Marchetti et al. 2003, 2008). MAD brilliantly demonstrated that MCAO works; however, the need for bright-enough NGS asterisms drastically limited the sky coverage to 50–100 scientifically interesting objects, almost exclusively close to the Galactic Plane.

This is easily understandable when one remembers that the sky coverage with a single NGS is approximately 1% at H band for a field of radius 20 arcsec—approximately the isoplanatic angle at 1.6 μm. MCAO requires at least three NGSs of brightness similar to the single NGS SCAO, within an FoV of a couple of arcminutes in diameter. Assuming a uniform distribution of star-apparent positions, this means a probability of $[0.01 \times (60 \text{ arcsec}/20 \text{ arcsec})^2]^3 \approx 0.003\%$. Given there are 47 million 2-arcmin-diameter fields in the 4-sr celestial sphere, this translates into a total of about 1,000 fields that are observable with NGS MCAO, which roughly matches the experience of MAD (out of these 1,000 fields, only a fraction will actually be of any scientific interest).

One way to significantly improve the sky coverage is to use LGSs. An arbitrary number of LGSs can be created (budget allowing) on which the high-order wavefront sensing can be done, as for instance seen in **Figure 7**. NGSs are still needed for the TT and plate scale mode sensing (see below), but because the full aperture can be used for these TT measurements, fainter guide stars can be used. A typical value for LGS SCAO sky coverage is 30% in the H band. Estimates calculated for GeMS also point to 30% H-band sky coverage. This is consistent with the first-order analysis presented above: $(0.3 \times 60 \text{ arcsec}/30 \text{ arcsec})^3 = 21\%$.

Alternative, more efficient NGS-based methods have been proposed by Ragazzoni et al. [2002, multiple field of view (MFoV) AO; 2014, global multiconjugate adaptive optics (GMCAO)], in which wider asterisms of bright NGSs are used for the ground-layer correction, whereas the

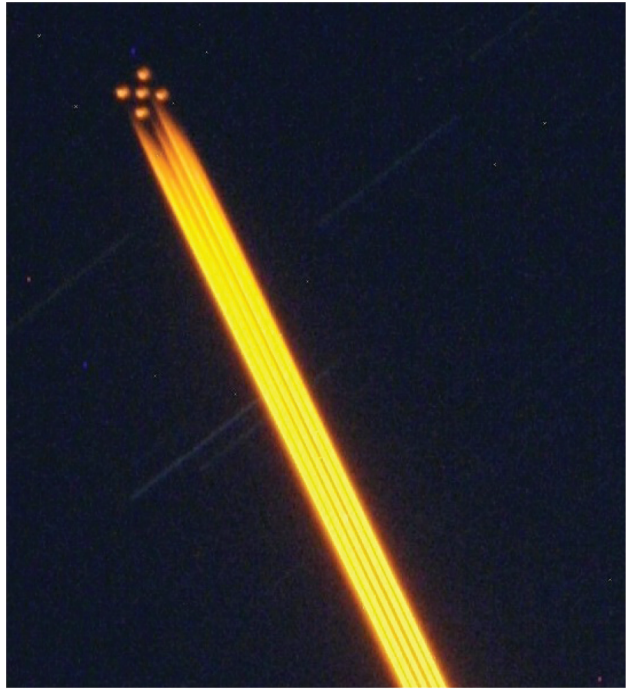

**Figure 7**

MCAO systems require multiple guide stars to reconstruct the full volume of turbulence. This image shows five laser beams used by GeMS, with Rayleigh scattering and the five-laser-guide-star asterism, a square with a central star, 60 arcsec on the side. Figure adapted from image provided by Maxime Boccas. Abbreviations: GeMS, Gemini MCAO System; MCAO, multiconjugate adaptive optics.

free atmosphere component—which is generally weaker than the ground component—can be sensed with WFSs having larger subapertures, therefore allowing the use of fainter NGSs, albeit in a narrower FoV[5]—hence the name. LINC-NIRVANA (see Section 3.4) is an MCAO system currently being commissioned on the LBT and is based on the MFoV concept (Farinato et al. 2008). It is worth noting at the threshold of the extremely large telescope era that the scaling laws of these NGS-based concepts are favorable to larger apertures (Viotto et al. 2015) for at least two reasons: First, assuming they work at or close to the diffraction limit of the telescope, some WFSs—like the pyramid WFS—become more and more sensitive as $D/r_0$ gets larger. The second reason is geometrical: The overlap of the beams coming from the stars of a given asterism becomes larger for larger apertures, making the tomography easier and more stable.

## 2.4. Tomographic Wavefront Reconstruction Methods

The tomographic wavefront reconstruction is the key step in providing the MCAO correction. Existing AO systems close the loop by driving the DM(s) in such a way as to cancel the signal on the WFS(s). This is accomplished using reconstructors that invert the interaction matrix, which describes the relationship between actuator commands and the WFS measurements. All current AO systems use these reconstructors.

---

[5]Whether using Shack–Hartmann or pyramid WFSs. In the latter the concept of subaperture has a natural analogy.

Relevant to tomographic systems, minimum variance reconstructors aim to optimize the correction over a number of points in the science field, rather than in the direction of the guide stars (Fusco et al. 2001, Ellerbroek 2002). The minimum variance approach splits the reconstruction process into two consecutive steps: The first is a tomographic reconstruction of the atmospheric layers on a set of predefined altitudes. The second is the projection (or least-square fit) of this tomographic information onto the DMs.

To improve performance, the tomographic reconstruction—or phase reconstruction step—makes use of a priori information on the averaged strength of the turbulent layers.[6] One issue is that in the classical closed-loop configuration (see Section 1.2), the WFS(s) is after the DM(s), and hence it only sees the residual turbulence, which does not obey the same statistics. In order to tackle this issue, Ellerbroek and colleagues (Ellerbroek & Rhoadarmer 1998, Ellerbroek & Vogel 2003) proposed an adaptation of the original algorithm, and among the potential solutions is the idea of reconstructing the virtual open loop measurements: a combination of the closed-loop measurement and a measurement computed from the current DM position. This method, called POLC for pseudo open-loop control, provides a stable and almost optimal MCAO performance (Gilles 2005, Piatrou & Gilles 2005). Note that the tomographic reconstruction theoretical development, taking into account the dynamical evolution of the turbulence, has been carried out by Le Roux et al. (2004). This method ensures the optimality of the reconstruction, at the price of a complexification of the real-time process (Gilles et al. 2003, 2013). Another field of study has also been devoted to the optimal data fusion between the NGS and LGS measurements when they are acquired at different frame rates (Correia et al. 2013). In this latter work, high-order correlations are used to better estimate the low-order modes (related to NGS) and potentially increase the sky coverage as fainter stars may be used.

The trade-off between optimal performance and real-time computation complexity has led to several developments, especially in the context of future extremely large telescopes for which the number of degrees of freedom becomes very large (a few hundred for 8-m telescopes versus tens of thousands for 38-m telescopes). Simplifications of the formalism have been attempted by means of a virtual DM approximation (Le Louarn & Hubin 2004), or a split tomography (Gilles & Ellerbroek 2008) that treats the measurements coming from LGSs and NGSs in a fully independent way. Complementary approaches take advantage of the sparsity of the matrices to develop conjugate gradient methods, as proposed by Thiébaut & Tallon (2010) or Gilles et al. (2003).

Finally, one can note that current operational MCAO systems have been using very basic, hence nonoptimal, tomographic approaches (Neichel et al. 2010, Quirós-Pacheco et al. 2010, Schmidt et al. 2016). This is mainly because of computing power limitations. Future extremely large telescope–MCAO systems are more ambitious and will take advantage of the latest developments in this field (Kerley et al. 2016).

### 2.5. MCAO-Specific Error Sources

MCAO—and LGS MCAO in particular—comes with its own error sources and limitations. The generalized fitting error, the tomographic error, and the generalized aliasing error are such MCAO limitations, whereas TT indetermination and fratricide are a consequence of the use of LGSs. A detailed description of these errors has been given by Ellerbroek (1994), Rigaut et al. (2000), Tokovinin et al. (2000), and Tokovinin & Viard (2001).

---

[6]As such, tomographic reconstruction can really benefit from online estimation of coarse $C_n^2$ profile, using, for example, SLODAR (slopes detection and ranging) or other methods (Cortés et al. 2012).

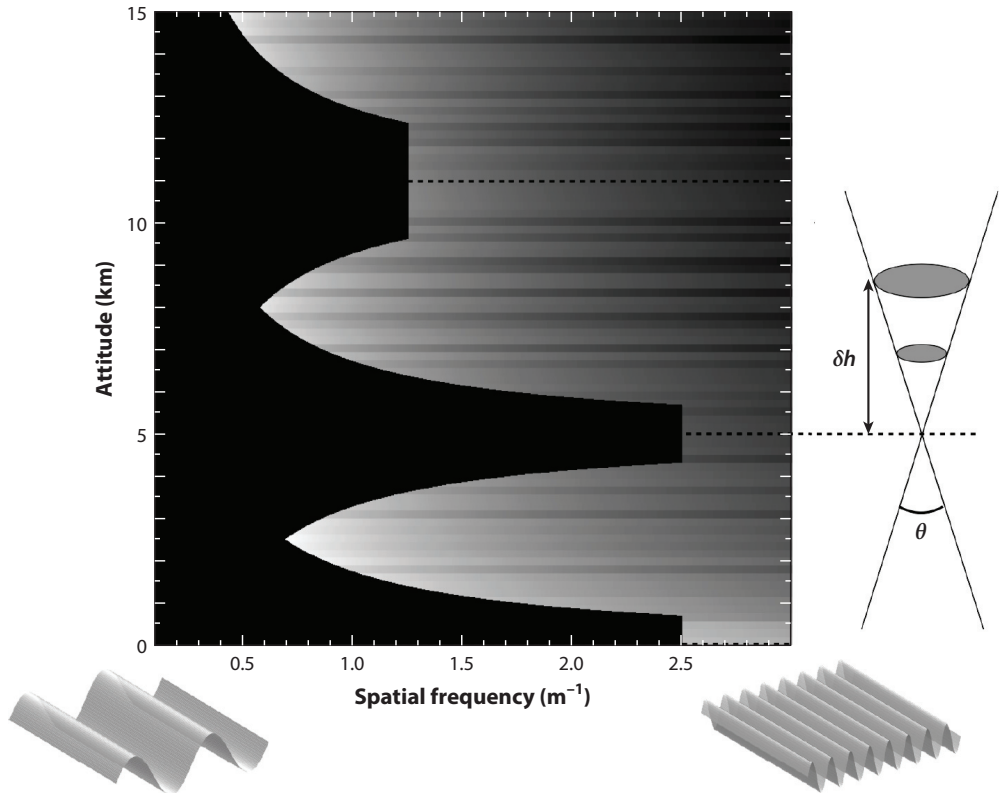

**Figure 8**

Illustration of generalized fitting, or the ability of an MCAO system to compensate for turbulence as a function of spatial frequency and altitude difference with respect to the nearest DM. This is an example of a toy system with a field of view of 120 arcsec and three DMs conjugated at 0, 5, and 11 km with actuator pitches of 20, 20, and 40 cm. The turbulence is stratified in layers, represented as horizontal lines; for each layer, the *x* axis shows the turbulence content versus spatial frequency (in effect the power spectrum), with gray level tracing the power at this particular frequency (*white*, most power; *black*, no power). The areas in black are the [spatial frequency, altitude] pairs that can be corrected by the DMs simultaneously for all directions in the field of view. Adapted from R. Ragazzoni, private communication. Abbreviations: DM, deformable mirror; MCAO, multiconjugate adaptive optics.

**2.5.1. Generalized fitting error.** This error results from the fact that there is a limited number of DMs available to fit the turbulence vertical $C_n^2$ profile.[7] This is illustrated in **Figure 8**.

Consider an MCAO system aiming to compensate for the turbulence-induced wavefront degradation over an FoV of $\theta$. By essence of the MCAO correction, in which we are seeking a uniform compensation over a finite sky area, this means that the corrections applied by a DM (e.g., the one at 5 km in **Figure 8**) have to be applied for all objects in the FoV. If a bump in the wavefront occurs at 5 km, it can be perfectly corrected by a counter-bump on the 5-km DM simultaneously for all directions in the FoV. But the ability of the same DM to correct for a bump occurring at a mismatched layer at altitude 5 km $+ \delta h$ is reduced because any figure on the DM is effectively blurred on the mismatched layer by a kernel of size $\theta \, \delta h$. In effect, this means that as the vertical distance between the DM and a turbulent layer increases, fewer and fewer high spatial frequencies can be corrected in the layer. Low spatial frequency modes will be well corrected, but for instance, a sine mode with a period smaller than ($\theta\ \delta h$) will not be correctable at all. The cutoff frequency can be approximated as ($\theta\ \delta h)^{-1}$.

---

[7]Hence the term generalized fitting error, as it can be viewed as a generalization of the classical AO fitting error due to the limited number of actuators in the aperture. Note that this error is called "generalised anisoplanatism error" by some authors, e.g., by Tokovinin et al. (2000), justified by the fact that it gets larger as the FoV increases.

The left-hand side of **Figure 8** gives a visualization of this effect. It illustrates the ability of an MCAO system to compensate for turbulence as a function of spatial frequency and altitude.

Several comments are warranted on this figure:

- Each DM correction domain extends up to a maximum spatial frequency that corresponds to the DM cutoff frequency $1/(2d)$, where $d$ is the actuator pitch. This is the usual AO fitting error.
- As said above, the outer envelope of each DM compensation area follows $\delta h = (f\theta)^{-1}$ ($f$ is the spatial frequency). Consequently, as $\theta$ increases or decreases, the dark bands become narrower or broader, respectively, leaving more or less turbulence uncorrected.
- It can readily be seen that in this particular example, a lot of turbulence is missed between the DMs. This would allegedly be fine if the turbulence profile was such that the turbulence layers were always at fixed altitude, but this is rarely the case. In reality, a system like this one would show poor performance overall and large performance variability. To show more resilience to $C_n^2$ profile fluctuations (not mentioning the fact that rarely is the turbulence limited to 2–3 layers), smaller $\theta$ have to be adopted, which in this figure will have the effect of filling the gaps between the DMs, illustrating the natural trade-off between FoV and generalized fitting error.

As a final remark, we note that both the vertical density of DMs as well as the FoV $\theta$ play roles in this error, justifying both the generalized fitting and generalized anisoplanatism terminologies.

**2.5.2. Tomographic error.** This error results from the improper sampling of some part of the turbulence volume (Gavel 2004, Le Roux et al. 2004, Quirós-Pacheco et al. 2010). Even when using multiple guide stars, some part of the turbulence volume above the telescope is only probed by one guide star. This necessarily limits the tomography, which will have no way to determine at which altitude the perturbation is located in this part of the beam. Another fundamental limitation of the tomography comes from unseen modes. The unseen modes are modes that, when combined at the DM altitudes and projected in the direction of the WFS, result in either phases that cancel out or phases to which the WFS is not sensitive. For instance, sine modes of the same spatial period, but of opposite amplitude, will cancel out when viewed by a WFS. It is possible for certain modes to achieve this condition for all MCAO sensors simultaneously, in which case the system will be blind to these modes, although they can significantly affect the image quality for objects in between the guide stars in the FoV. Both these effects have been described by Rigaut et al. (2000), Le Louarn & Tallon (2002), Le Roux et al. (2004), and Neichel et al. (2008).

**2.5.3. Generalized aliasing error.** In classical AO aliasing, a high-order aberration is seen by a WFS as a low-order aberration. In MCAO, turbulent layers above and below the altitude control domain of the MCAO system (e.g., any layer above the highest conjugated DM) will be seen and wrongly interpreted as layers inside the control domain, which is the only thing the system has knowledge of. This creates an additional error called a generalized aliasing error, because this can be seen as a generalization of the AO aliasing process. Unlike classical spatial aliasing, which is a property of the WFS, the generalized aliasing is a property of the tomographic (or more generally reconstruction) process. Some reconstruction processes perform better than others to mitigate this effect (Quirós-Pacheco et al. 2010).

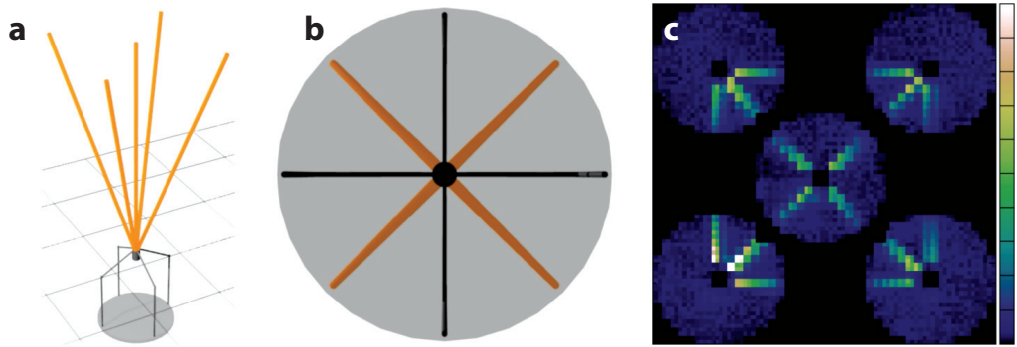

**Figure 9**

Illustration of the fratricide effect in multi-LGS systems. (*a*) Configuration of the LGS beams for GeMS. (*b*) Same beams, as seen by the WFS staring at the central LGS. (*c*) Actual image of the subapertures' total intensity of the five GeMS WFSs, showing the impact of the fratricide effect. The laser has been detuned off the sodium D2 line, so that the LGSs disappear, leaving only the Rayleigh scattering. Abbreviations: GeMS, Gemini MCAO System; LGS, laser guide star; WFS, wavefront sensor.

**2.5.4. Tip-tilt and plate scale control.** This is the generalization to MCAO of the TT indetermination problem described in Section 1.4.2. LGSAO only has an issue with TT sensing. In MCAO, not only the average TT over the field but also the differential TT between the guide stars has to be corrected, which is nothing else than a dynamic variation of plate scale across the FoV (Flicker et al. 2003). Several methods have been proposed to solve this (Ellerbroek & Rigaut 2001, Gilles & Ellerbroek 2008); some of them entail the use of multiple TT NGSs, others a mix of natural, Rayleigh, and sodium guide stars (Le Louarn 2002).

**2.5.5. Fratricide.** Before creating the sodium LGS at 90 km altitude, the laser light is scattered by low-altitude air molecules, creating the Rayleigh plume. This makes for beautiful photos of beams crisscrossing the night sky; however, this creates a nuisance for other telescopes[8] and can affect the telescope that launches the laser itself as well. Rayleigh scattering from the laser can find its way to the science instrument or can pollute other LGS WFS signals. The latter can occur depending on the laser launch configuration; it will necessarily occur when two or more lasers are launched from the same point, i.e., the same laser launch telescope (LLT). This is referred to as fratricide, as it traces a nefarious interaction between two or more lasers. **Figure 9** illustrates this effect.

Several ways have been proposed to mitigate fratricide:

- The most obvious and efficient way is to use side-launch, in which the various lasers are launched using LLTs physically located outside the envelope of the telescope primary mirror. This configuration, when carefully crafted, can prevent fratricide altogether. The flip side is that it will increase LGS elongation and possibly increase hardware cost as one LLT per laser is required.
- Use a Rayleigh LGS, with synchronization of the multiple laser pulses. One significant issue with Rayleigh guide stars is the large cone effect that results from the low altitude of the created guide star (typically 15–20 km). This makes for less effective coverage of the turbulence volume, often meaning more stars are needed. Note that this is a nonissue for,

---

[8]This is usually managed at the observatory level by the Laser Traffic Control System, which monitors and notifies parties about potential beam collisions.

- e.g., GLAO for which only the measurements of the ground layer are needed, as beautifully demonstrated by ARGOS (Abundances and Radial Velocity Galactic Origins Survey; Rabien et al. 2017).
- Calibrate and subtract the fratricide as an offset in the WFS data. This is theoretically possible, although it will add some photon noise to the measurements (for GeMS, the Rayleigh is typically up to 10 times the useful LGS signal in affected subapertures). However, this requires a stable beam, which is not the case in GeMS, and has therefore not been implemented (Neichel & Rigaut 2011). NFIRAOS (Narrow Field Infra-Red AO System) at Thirty Meter Telescope (TMT) is planning to use this method.
- Use a pulsed sodium GSL (Holzlöhner et al. 2012, Rochester et al. 2012). This has been the Holy Grail of GSL technologies, as it would provide a solution not only to the fratricide but also to the elongation problem, which is seriously impacting AO on extremely large telescopes. However, the required small duty cycle this necessitates would significantly impact the sodium return. The laser development challenges are also significant, which might explain why this has never really taken traction to date.

Note that the impact of fratricide is proportionally less severe for an extremely large telescope, because a smaller fraction of the radius of the telescope will be affected by it, the effect being proportional to the absolute distance from the launch to a subaperture (assuming the same angular difference between the LGSs). **Figure 9c** shows that at Gemini, the Rayleigh almost dies off for subapertures 4 m away from the launch (one telescope radius). On a 30-m telescope, this would represent $4/15 \approx 25\%$ of the radius. This is partly why the TMT has chosen to maintain a center-launch of its six-LGS asterism.

## 3. REAL-WORLD SYSTEMS

This section describes actual MCAO systems, and discusses practical limitations and trade-offs that have to be made during their design.

### 3.1. Practical Limitations and Design Considerations

With MCAO systems being more complex than SCAO, they face additional constraints and challenges. On top of the fratricide effect discussed in Section 2.5.5, a nonexhaustive list includes the following:

- For the LGS-based systems, the LGS facility is more complicated than it is for a single-beam facility. The laser has to be more powerful, of course, and either a single beam has to be split (e.g., GeMS) or multiple beams have to be fed to a single LLT, or multiple beams and multiple LLTs have to be used (e.g., ESO AOF). On Alt-Azimuth-mount telescopes, the rotation of the LGS asterism has to be addressed: The asterism can either corotate with the science field to maintain the LGSs fixed with respect to it (and thus ensure more constant performance across the field) or stay fixed with respect to the telescope.
- Another complication of using LGSs is that the distance from the telescope to the sodium layer varies with the telescope elevation. In a single LGS system, this is generally compensated for by a focus mechanism, either the entire WFS moving back and forth or some kind of optical focus compensation. In MCAO, there are several guide stars. If these are processed by a single, multi-LGS WFS, the latter will have to preserve good optical quality for on- and off-axis objects over the whole possible LGS range. This makes for challenging optical design.

- Some early MCAO designs considered putting the DMs on rails, so that their conjugation altitude could be adapted to the current condition to minimize generalized fitting and optimize performance. This has only been adopted by solar MCAO (Schmidt et al. 2017), made possible by the convenience of a Coudé lab. It is also debatable whether this is beneficial given the spread and the changing nature of the vertical turbulence profile.
- The order (ground then altitude, or the opposite) in which the DMs correct the beam is theoretically relevant because phase corrections propagate into amplitude fluctuations. Flicker (2001) and Roggemann & Lee (1998) found that this effect is negligible except for cases of strong turbulence like horizontal propagation (e.g., ground to ground mostly for military applications) and perhaps also for solar MCAO, as turbulence conditions are more severe than for nighttime astronomy.
- On GeMS, and to a lesser extent on MAD, the high-altitude DM has a lower actuator density than the low-altitude one(s). This is motivated by the fact that because of the FoV, the physical surface $S$ to be corrected increases with altitude, following $S(h) = \pi(R + \theta h)^2$, $R$ being the telescope radius, $\theta$ the half FoV, and $h$ the conjugation altitude. In the case of GeMS, the surface to be controlled almost doubles from the 0- to 5-km DM ($\approx$50 to 90 m$^2$) and almost triples for the 9-km conjugation (140 m$^2$). Because the cost of DMs is essentially proportional to the number of actuators, that would mean that 50% of the DM budget would go to the high-altitude DM if the same actuator density was used (in terms of actuators per square meter as projected on the altitude layer). Yet, the high-altitude DM only compensates for typically 10–20% of the total turbulence. Fortunately, because the turbulence carried by the high-altitude layer is weak, the associated $r_0$ is large, and thus a lower actuator density can be used. This is also behind the concept of GMCAO (see Section 2.3).
- An obvious trade-off, fed by the science requirements and addressed via simulations using extensive site monitoring information ($C_n^2$ profile, etc.) is between FoV, Strehl ratio, and PSF uniformity: Because of generalized fitting, Strehl will decrease when increasing the FoV. Because of tomographic errors, the performance will be less and less uniform as the FoV increases.
- Another trade-off is in how many LGSs and NGSs are used. More LGSs mean a better and more uniform performance. Many studies have been carried out, and they indicate that for FoVs of 40–60 arcsec, in the NIR and for 8-m telescopes, performance does not much increase after six LGSs (e.g., Le Louarn et al. 2006). The number of guide stars has only a weak dependence on the telescope diameter but a relatively strong one on the FoV and compensation wavelength.

In the following sections, we examine in a little more detail the four MCAO systems that have seen starlight: MAD; GeMS; LINC-NIRVANA; and the solar-AO system.

### 3.2. Multiconjugate Adaptive Optics Demonstrator

The MAD project started roughly at the same time as GeMS but saw first light much earlier. MAD was built as a demonstrator, to prove the validity of the MCAO concept and to probe its performance and limits (Marchetti et al. 2003). It sprung from the success of the first on-sky active compensation tomographic experiment (Ragazzoni et al. 2000b), in which the phase measured on three off-axis NGSs was used to compensate in a fourth direction (another star at the approximate center of the FoV).

Part of MAD involved testing two radically different MCAO wavefront sensing modes: star-oriented and layer-oriented, as they were dubbed by the MAD team (or perhaps by Roberto Ragazzoni). Both modes are purely NGS based.

> **PYRAMID WAVEFRONT SENSORS**
>
> In these WFSs, a pyramid (four-facet prism) is placed at the telescope focus; a set of postfocal optics then form four pupil images on a detector. The ratio of intensities in pixels corresponding to the same region of the pupil gives a direct measurement of the *X* and *Y* slopes of the wavefront at this pupil location. Pyramid WFSs offer improved performance compared with traditional techniques like Shack–Hartmann WFSs, because their responsiveness increases as the image quality gets better (as one gets closer to $\lambda/D$), improving the overall sensitivity of the system by a factor of two or more.

In the star-oriented mode, $N$ WFSs are looking at $N$ guide stars; each WFS measures the phase integrated along the line of sight in the column of atmosphere above the telescope; all those integrated phase information are fed to a tomographic reconstruction process that evaluates the turbulence volume and computes the command to apply to each DM to optimize the compensation in the FoV, according to some criteria. The star-oriented mode used three $8 \times 8$ Shack–Hartmann WFSs and two DMs (both 60-actuator curvature DMs) conjugated at 0 and 8.5 km above ground.

In the layer-oriented mode, there is one WFS per layer. The layer-oriented WFS is based on a pyramid WFS (see the sidebar titled Pyramid Wavefront Sensors) (Ragazzoni 1996, Esposito & Riccardi 2001), but is extended to work with multiple guide stars at the input. Using a set of front optics, each NGS is optically coadded on a single detector, conjugated at the altitude at which the DM is to be placed. The light is split into as many channels/detectors/altitudes as there are DMs. It can be argued that this method uses photons more efficiently than the star-oriented method. Potentially, many guide stars can be used in the FoV (in the MAD case, up to eight), improving the limiting magnitude of the system. The layer-oriented approach is also at the heart of the MFoV concept (see Section 3.4), which provides additional advantages in terms of computational processing needs and sky coverage. A detailed description of the layer-oriented concept is given by Ragazzoni et al. (2000a).

MAD was a project led and mostly developed at ESO, with the collaboration of Universidade de Lisboa and observatories of Padova and Arcetri (INAF, Italian National Institute for Astrophysics). It was mounted at the Nasmyth platform of the VLT UT3 telescope, and used over several semesters for technical tests and astronomical programs. Typical results (in star-oriented mode) show large gains over an FoV of up to 2 arcmin. FWHM reduced from the regular seeing (0.7–1.2 arcsec) down to typically 100 mas; Strehl increased up to 40% in the $K_s$ band (Marchetti et al. 2007, 2008), as discussed in Section 4 and shown in **Figure 10**.

### 3.3. Gemini MCAO System

GeMS is an MCAO system in use at the Gemini South telescope. It uses five LGSs feeding five $16 \times 16$ Shack–Hartmann WFSs and needs three NGSs and associated NGS WFSs (three TT WFSs and one focus WFS) to drive two DMs. A full description is given by Rigaut et al. (2014). It delivers a uniform, close to diffraction-limited NIR image over an extended FoV of 2 arcmin$^2$. GeMS is a facility instrument for the Gemini South Chile telescope and, as such, is available for use by the extensive Gemini international community. It has been designed to feed two science instruments: Gemini South Adaptive Optics Imager (GSAOI) (McGregor et al. 2004), a 4k × 4k NIR imager covering $85 \times 85$ arcsec$^2$, and Flamingos-2 (Elston et al. 2003), an NIR multiobject spectrograph. The GeMS project started in 1999 with a conceptual design study. Design and

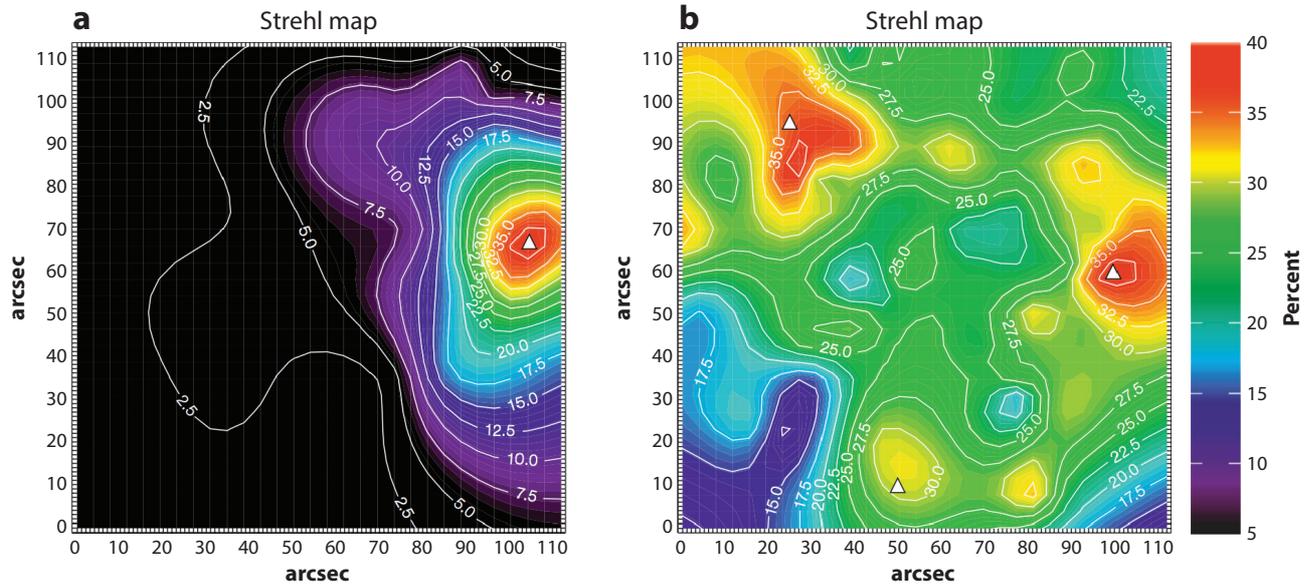

**Figure 10**

Strehl ratio performance gain from (*a*) SCAO to (*b*) MCAO with MAD in star-oriented mode. The Strehl ratio contours are labeled in percentages. The NGS locations are marked by the white triangles. Figure adapted from Marchetti et al. (2007). Abbreviations: MAD, Multiconjugate Adaptive Optics Demonstrator; MCAO, multiconjugate adaptive optics; NGS, natural guide star; SCAO, single conjugate adaptive optics.

construction lasted approximately 10 years. On-sky commissioning started in January 2011 and culminated in December 2011 with images having an FWHM of $80 \pm 2$ mas at 1.65 μm (H band) over the entire 85-arcsec GSAOI FoV.

GeMS works with an asterism of five LGSs, with four of the LGS spots at the corners of a 60-arcsec square and the fifth located at the center. These LGSs are produced by a single 50-W laser split into five distinct 10-W beacons. The on-sky performance of the Laser Guide Star Facility is described by d'Orgeville et al. (2012). GeMS was originally designed to work with three DMs conjugated at 0-, 4-, and 9-km altitude. Following issues with one of the DMs during early commissioning, it has since then operated with the 0- and 9-km DMs only, making the generalized fitting more severe and a major factor limiting its performance. GeMS has over 20 loops, offloads or supervisory processes, the calibrations of which required over 30 nights of commissioning time (Neichel et al. 2014c).

### 3.4. LINC-NIRVANA

LINC-NIRVANA shares a lot of its DNA with MAD. It is currently being commissioned on the LBT (Kopon et al. 2014). LINC-NIRVANA is essentially a Fizeau interferometer combining the two LBT 8.4-m beams, each of which is corrected by a layer-oriented MCAO system using NGSs. It uses the telescope adaptive secondary mirror for ground-layer correction and two Xinetics 349-actuator DMs for the high layer (one per beam, conjugated at 7.1 km). The wavefront sensing follows the MFoV approach (Farinato et al. 2008), in which the low-altitude-layer WFS covers a larger area of the Xinetics 349-actuator DMs than the high-altitude-layer WFS (essentially because the beams stay overlapped over considerably wider field angles for the ground layer than for the high-altitude layer). In LINC-NIRVANA, the ground-layer DM receives input from a 6-arcmin field (up to 12 NGSs), whereas the high-layer loop samples a 2-arcmin field (up to 8 NGSs). LINC-NIRVANA has already demonstrated GLAO mode (Kopon et al. 2014), with 2× FWHM improvements.

### 3.5. Solar MCAO

MCAO, as a concept, is intimately linked to solar AO. After all, five years after first proposing the concept of MCAO (Beckers 1988), Jacques Beckers went on to take the directorship of the National Solar Observatory. AO has always been a very active field in solar astronomy, with successes being reported as early as the end of the 1980s (von der Luehe 1991, Rimmele 2004, Rimmele et al. 2013).

There are some fundamental differences between nighttime and daytime AO: First off, the turbulence is almost always more severe during the day; but the telescopes are also smaller, meaning the $D/r_0$ ratio is comparable to the one of nighttime AO systems on 8-m telescopes. Second, there is no single, isolated guide star on the solar surface. Instead, solar WFSs[9] have to use the moderately contrasted Sun photosphere as a guide object and use correlators to measure subaperture-to-subaperture position offsets. Of course, correlations need images with a nonzero FoV, which can often be larger than the small isoplanatic angle (solar-AO systems tend to work in the visible or red, which makes for small $\theta_0$ under the strong turbulence conditions encountered during the day), leading to some kind of SNR/anisoplanatism compromise and trade-off. Initial MCAO efforts met some success (Berkefeld et al. 2010), but the big breakthrough happened in 2016 at the Big Bear Solar Observatory on the solar-AO system, Clear, as reported by Schmidt et al. (2017) and illustrated in **Figure 11**.

## 4. PERFORMANCE

### 4.1. Overview of Current Performance

An example of typical performance obtained with current MCAO systems is illustrated in **Figure 10**, where the two panels show the measured Strehl ratio obtained from SCAO (panel *a*) and MCAO (panel *b*) with MAD. Conversely, the GeMS performance has been analyzed in further detail by Vidal et al. (2013), Neichel et al. (2014c), and Sivo et al. (2017) and is reported in **Figure 12**, which shows the delivered Strehl ratios and FWHMs measured under different seeing conditions. Those last results are based on images with exposure times between 10 and 180 s. There are, respectively, 950 points for the K-band images, 454 points for the H-band images, and 243 points for the J-band images. Additional performance characterization is given by Saracino et al. (2015), Turri et al. (2015), Dalessandro et al. (2016), Fiorentino et al. (2016), Massari et al. (2016a), and Bernard et al. (2016), with results that are consistent with those in **Figure 12**. Finally, performance was also evaluated at wavelengths closer to the visible (Hibon et al. 2014).

The GeMS performance in terms of Strehl ratio falls short of the original specifications by almost a factor of two. The reasons for this are well understood (Neichel et al. 2014d) and fall in three different categories:

- The failure of one (out of three) DMs, increasing the generalized fitting error (see Section 2.5.1);
- A lower-than-anticipated return flux from the LGSs, increasing the noise and the servo lag errors (see Section 1.3), and finally; and
- A poor optical throughput of the tip-tilt-focus WFS, increasing the residual tip-tilt.

All three items are being addressed by an upgrade plan for GeMS.

---

[9]Wavefront sensing on the Sun generally uses Sun spots, or the granules of faint contrast, as guide objects.

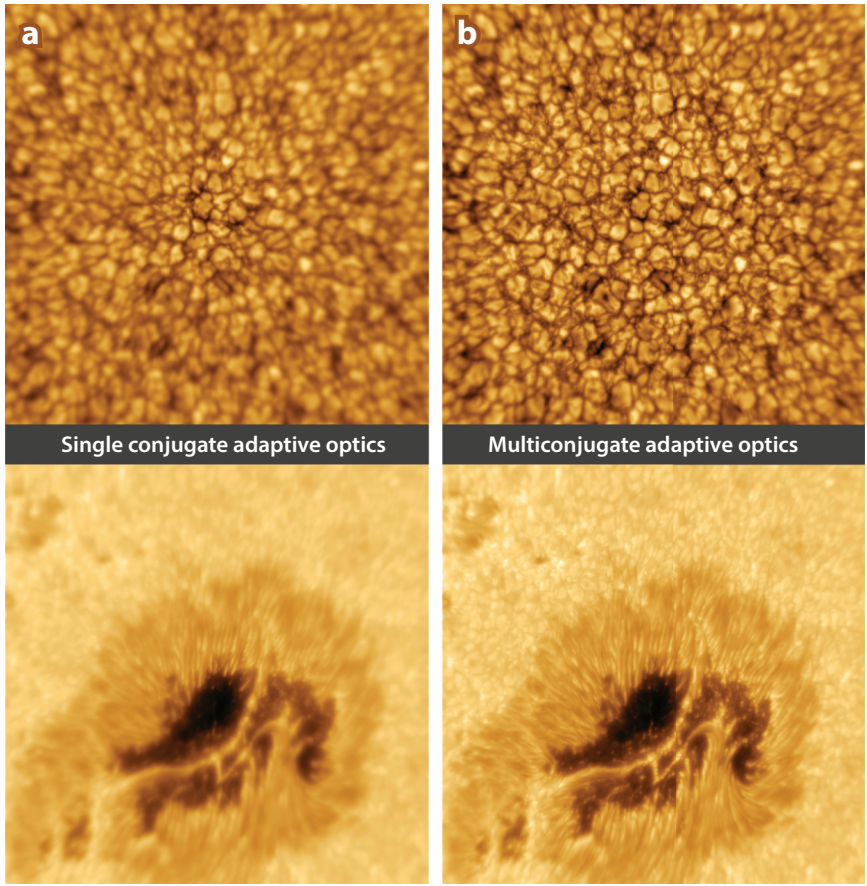

**Figure 11**

The first successful MCAO solar observations with the solar-AO system, Clear, at the Big Bear Solar Observatory (Schmidt et al. 2017). The field of view is 53 × 53 arcsec$^2$. (*a*) Images taken with SCAO, showing a clear anisoplanatic PSF degradation. (*b*) Images taken with the same system in MCAO mode, significantly enlarging the corrected field of view. The system is using three DMs conjugated at 0, 3, and 8 km for these particular images. Bottom images are taken at 705.7 nm; top images at 430.5 nm. Abbreviations: DM, deformable mirror; MCAO, multiconjugate adaptive optics; PSF, point spread function; SCAO, single conjugate adaptive optics.

In terms of PSF uniformity across the corrected field, and still based on GeMS experience, the relative rms variation of the FWHM across the images is of the order of 4% over a field of 1 arcmin$^2$, with the peak-to-peak variation being of the order of 12% of the average FWHM. A detailed analysis of the PSF shape has also been performed by Turri et al. (2017) and Dalessandro et al. (2016) for GeMS observations on crowded fields. The final performance obviously depends on the guide star geometry and atmospheric conditions, but it typically scales with the seeing.

### 4.2. Photometry

Under median seeing conditions and for isolated point sources, MCAO brings a 1.5- to 1.7-mag sensitivity gain with respect to seeing-limited imaging over the 1–2.5-μm range. For crowded fields, this gain may go up to 3 mag. This gain in sensitivity may, however, be affected by systematic photometric errors. The delivered photometric accuracy directly depends on the PSF

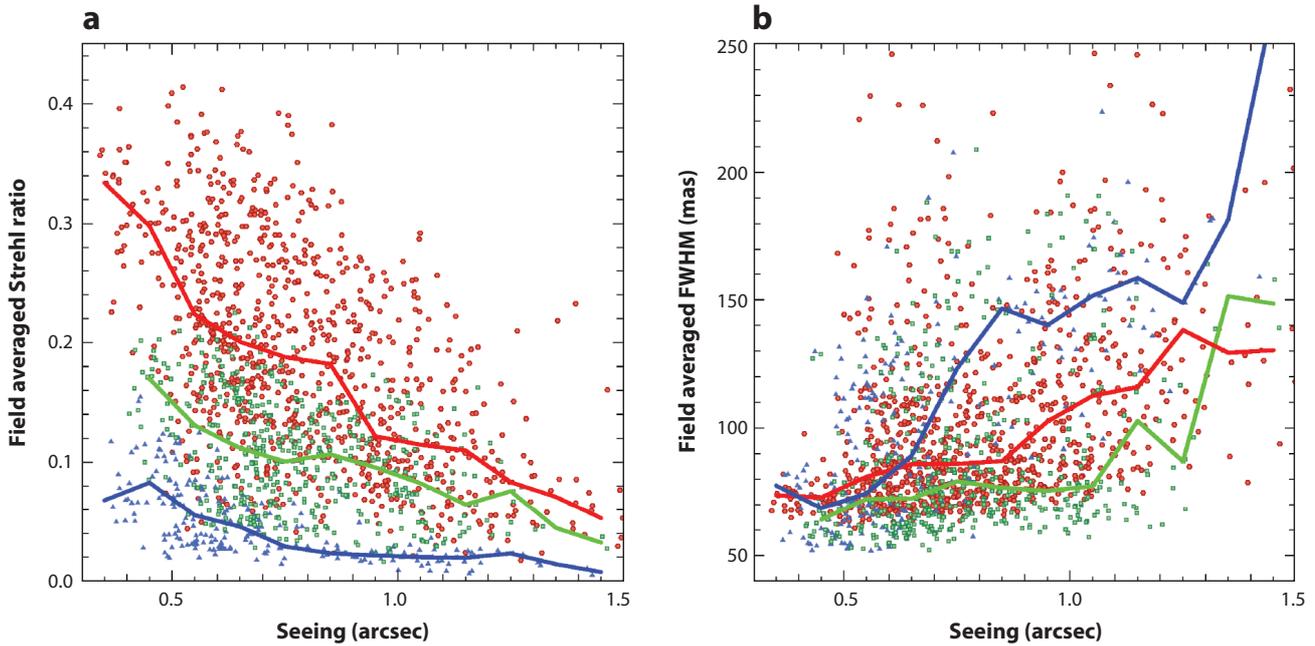

**Figure 12**

Summary of (*a*) the Strehl ratio and (*b*) FWHM obtained for K-band images (*red*), H-band images (*green*), and J-band images (*blue*) with GeMS, based on commissioning and science verification data. The median natural seeing over these observations is 730 mas (defined at 0.5 μm). For reference, the diffraction-limited FWHMs are, respectively, 32, 43, and 55 mas for J-, H-, and K-band filters. In terms of image quality for a given fraction of the observing time, GeMS delivers a FWHM of 75 mas (or better) in the H band for 50% of the time. Adapted from Neichel et al. (2014c). Abbreviations: FWHM, full width at half maximum; GeMS, Gemini MCAO System.

uniformity and stability across the MCAO corrected FoV and over time. As shown by **Figure 10**, the PSF uniformity provided by an MCAO system is significantly improved compared with SCAO. However, the PSF shape may present residual ellipticity that has to be taken into account during the photometry and astrometry analysis process in order to get the best scientific use of the data. Turri et al. (2017) performed a detailed frame-by-frame study of PSF variations over the field. In particular, they found that the PSF ellipticity may change from one exposure to the next. The complex PSF shape then calls for specific and dedicated photometric tools, allowing for a possibly complex spatial variability and the processing of the exposures independently (Ascenso et al. 2015). Another aspect concerns the photometric calibration (zero point) and cross-calibration with external catalogs. Most current catalogs have been built from seeing-limited (or low angular-resolution space) observations, and blending in reference catalogs must be handled carefully. This can be done by including unresolved companions or selecting only isolated stars in the zero-point determination. Applying such a dedicated photometric analysis, Turri et al. (2017), Saracino et al. (2016), and Massari et al. (2016a) demonstrated a total photometric accuracy between 0.01 and 0.05 mag for the brightest stars (J < 18), and better than 0.2 mag for limiting magnitudes of 23 in the J band.

### 4.3. Astrometry

The astrometry gain provided by MCAO compared with seeing-limited observations can easily reach an order of magnitude (see Section 2.2), but MCAO may also introduce new systematic errors that would prevent this gain from being achieved. A first astrometric error budget for

*Rigaut • Neichel*

existing MCAO systems is described by Meyer et al. (2011), Lu et al. (2014), and Neichel et al. (2014b). The classical astrometric error budget (Fritz et al. 2010, Trippe et al. 2010) includes instrumental effects (e.g., geometric distortion), atmospheric effects (e.g., chromatic differential refraction), and astronomical uncertainties (reference source selection). Out of the different terms presented, the quasi-static distortion variations have been identified as the main offenders for GeMS astrometric performance. As such, Neichel et al. (2014b) concluded that for single-epoch data sets, an astrometric error of approximately 150 µas can be reached, but for multi-epoch data sets, a systematic noise floor of ∼400 µas was limiting the final performance. A precise determination of the static distortion has been carried out by Massari et al. (2016b) and Dalessandro et al. (2016) down to submilliarcsecond accuracy. Unfortunately, it appeared that this distortion solution was not static and will evolve over different observation periods (Bernard et al. 2018). As a lesson learned from those first MCAO systems, the ultimate astrometry performance requires dedicated effort and instrument design. In particular, it may not be possible to get an absolute and stable enough instrument; hence having ways to calibrate and/or postprocess the data is mandatory. For instance, one strategy that has been developed for GeMS was to minimize the dithering pattern during the observations and reproduce the star position on the science detector from one epoch to the next. For GeMS and MAD, the astrometric performance was not defined as a scientific requirement, and those instruments are missing dedicated astrometric calibration tools. Future instruments, and especially those for the extremely large telescopes, will incorporate calibration, observation, and data reduction processes dedicated to astrometry (Rodeghiero et al. 2016, Schöck et al. 2016).

**Phase diversity:** a method allowing retrieval of phase information based on focal plane images, often used in AO to measure the noncommon path aberrations

### 4.4. Noncommon Path Aberrations

AO compensates for dynamic atmospheric turbulence aberrations. It also compensates for static optical aberrations in the optical system. Obviously, it only does so for the aberrations that are seen by the WFSs. Generally, in an AO system, part of the light is intercepted at some location in the optical system and sent to the WFS. The split can be done by a beam splitter (amplitude split), a dichroic (wavelength split), or a pick-off mirror (spatial split). If some aberrations are present downstream of the split in the science optics, they will not be seen by the WFS and therefore will not be corrected. Conversely, if some aberrations are present in the WFS optics, they will be seen; thus they will be compensated for and applied to the science image even though they should not be. These aberrations are called noncommon path aberrations (NCPAs). The usual method to compensate for them is to change the zero point of the WFS so that the loop will aim to converge not to a zero WFS signal but to some static point that has been calibrated to correspond to the desired optical aberrations.[10] In this way, one can either subtract WFS path aberrations or add science path aberrations. This calibration is usually done off-line using a calibration source and adopting some kind of metric on the science image itself; that can be "power in the bucket," or more elaborate methods like focal plane wavefront sensing (Korkiakoski et al. 2014), of which phase diversity is a prime example (Blanc et al. 2003). These methods are generally iterative and result in static PSFs with Strehl ratios in excess of 90% (Sauvage et al. 2007, Ren et al. 2012, Antonello & Verhaegen 2015).

The problem becomes more complicated with MCAO systems: The PSF has to be optimized everywhere in the FoV of the science instrument, simultaneously. The problem boils down to finding the shape to apply to the DMs to optimize the image quality over the FoV. This was first described by Kolb (2006), who proposed a fast method to compensate for MAD NCPAs. For GeMS, Gratadour & Rigaut (2011) proposed an iterative tomographic approach in which phase diversity images are used to estimate the best DM commands to cancel these aberrations (over all DMs), obtaining static Strehl ratio of approximately 90% in H band. In a way reminiscent of the generalized fitting error, note that if a static aberration comes from an optical element conjugated to an altitude outside of the range addressable by the DMs, it cannot be corrected; this is an important point to consider for the design of future MCAO systems.

---

[10]In view of the MCAO NCPA discussions (see the next paragraph), it is important to realize that it is the WFS(s) that sets the amount of NCPAs, but it is the DM(s) that actually corrects for it.

Another important aspect of NCPA is that any drift in these aberrations or in the LGS WFS calibrations will lead to static or quasi-static shapes on the output science images. Such static shapes are often seen on the science PSFs and are considered in the astrometric and photometric error budget (Neichel et al. 2014b, Turri et al. 2017). It is then of prime importance to try to reduce the absolute amount of NCPAs in the optical design, or even better, to implement a low-order WFS near the science focal plane (a.k.a. truth sensor) that would measure those quasi-static aberrations and send this information back to the main loop to compensate for it.

## 5. THE FUTURE

After the MAD (Marchetti et al. 2008) and GeMS (Neichel et al. 2014c, Rigaut et al. 2014) demonstrated in the 2000s that near-diffraction-limited performance could be obtained with uniform PSFs over extended FoVs, we now see the emergence of a second generation of MCAO systems, for the extremely large telescopes, on 8-m telescopes pushing toward the visible, and on the 4-m Daniel K. Inouye Solar Telescope (DKIST).

### 5.1. MCAO for Extremely Large Telescopes

Understandably, the extremely large telescopes heavily rely on AO for their instrumentation. From their inception, it was clear that these giants would not be built just to improve photon collection but that the gain in angular resolution was paramount in achieving their ambitious main science goals. Notwithstanding, some kind of active control helps tremendously to stabilize what has to be—by virtue of their sheer size—intrinsically floppy structures. To date, out of the three extremely large telescope projects, two have MCAO as part of their first generation instruments: The ESO ELT has MAORY (Multi-conjugate Adaptive Optics Relay), and the TMT has NFIRAOS.[11]

MAORY (Diolaiti et al. 2017) will feed MICADO (Multi-AO Imaging Camera for Deep Observations), an NIR camera and spectrograph. It is currently in ESO's phase B. It will use the ESO ELT M4 DM, with the supplement of one or two postfocal DMs, six LGSs, and three NGS WFSs (à la GeMS).

NFIRAOS (Herriot et al. 2014) will use two postfocal DMs, at 0 and 11.2 km, to provide near-diffraction-limited performance over the central 10–30 arcsec FoV. It will use six LGSs and three on-instrument WFSs for TT information on NGSs. A particularity of NFIRAOS is that it is contained in an enclosure that will be cooled to −30°C, which will minimize thermal emissivity from the many optics. Another notable feature is that it uses a four off-axis parabolas optical train to cancel out the large field distortion that a two-parabola system introduces (lesson learned from GeMS).

---

[11]The Giant Magellan Telescope has LTAO and GLAO, but no MCAO as part of the first-generation suite.

Both systems are targeting 50% sky coverage or more, with a uniform Strehl ratio of 30–50% at K bands. Because of the sheer size of the telescope apertures, combined with the large FoV that has to be transferred by the optics, both MAORY and NFIRAOS are gigantic instruments: 10.5 m × 8 m × 4.5 m for NFIRAOS and about 7 m × 7 m × 5 m for MAORY. One important characteristic of these new-generation AO systems is that, for the first time, they are designed alongside the telescope. The latter relies on the AO almost as much as the AO relies on the telescope.

### 5.2. Solar and Visible MCAO

Solar astronomers are planning to do an MCAO upgrade to the AO system currently built for the 4-m DKIST in Maui (Moretto et al. 2004, Johnson et al. 2016). The system currently has a 1,600-actuator ground-conjugated DM. The MCAO upgrade is in the preliminary design phase, with Clear as a pathfinder (Schmidt et al. 2017). Provisionally, it includes three altitude-conjugated DMs (with options for more), and a Shack–Hartmann WFS with 32 × 32 subapertures. Each subaperture has a large FoV in which a grid of 3 × 3 smaller areas are extracted and used as input for the correlator. In effect, this provides nine different WFSs to feed the tomography. The FoV is still being worked on, but initial simulations seem to indicate that a corrected field of 20–40 arcsec is possible at close to 30-mas angular resolution.

More recently, a visible MCAO system upgrade has been proposed for the ESO UT-4 AOF (Esposito et al. 2016). Targeting wavelengths down to 500 nm, results from initial simulations indicate that an angular resolution of 20 mas or better could be achieved uniformly over an FoV of approximately 30 arcsec. Such an upgrade for the ESO UT-4 AOF, even though challenging in terms of AO technology, would address a unique scientific niche in a post–HST era and would represent a perfect complement to extremely large telescope NIR capabilities. Outside of astronomy, ophthalmology has been using AO since the 1990s and recently started experimenting with MCAO concepts (Thaung et al. 2009).

### 6. CONCLUSION

In this review, after a brief reminder of the basics of AO, we have presented MCAO: what it is; how it helps or enables some astronomical programs; and its principles, limitations, specific error sources and tomographic reconstruction methods. Through real-world examples, we have tried to explain the practical aspects of the technique and what special design considerations are needed. Finally, we looked at MCAO performance in terms of PSF uniformity and photometric and astrometric performance.

Even though MCAO has been demonstrated, it is still the subject of active research and development. New advanced concepts like MFoV are promising. What could really propel the next quantum step in MCAO is the advent of true continuous 3D achromatic phase correctors: These would almost necessarily be transmissive, providing huge advantages for optomechanical packaging. By allowing continuous sampling and correction of the turbulence volume above the telescope, these yet-to-exist true 3D correctors would also provide a solution to generalized fitting; with them, it would be possible to further enlarge the FoV accessible with MCAO to several arcminutes or more.

After a slow start, perhaps due to the disbelief of the community that such a complex technique could be put to work, the interest of the community in MCAO is growing. MAD at ESO, GeMS at Gemini and Clear at the Big Bear Solar Observatory have demonstrated that MCAO works and provides uniform, almost diffraction-limited images over fields of 1 arcmin$^2$ or more. This is 10–20× larger in area than previous classical SCAO. Five new major instruments are currently in the works for the LBT, DKIST, VLT, TMT, and ESO ELT. The considerable gains brought by MCAO, coupled with the power of the extremely large telescopes, should make for a powerful combination. Furthermore, a visible MCAO system on the VLT—with a well-selected suite of instruments—could have a tremendous impact in the post-HST era.

**SUMMARY POINTS**

1. MCAO was proposed by Jacques Beckers in 1988 as a novel AO concept that allows for considerably widening the corrected FoV compared with SCAO systems—for instance, the Gemini MCAO system GeMS provides a gain of up to 20 in the AO-compensated area compared to the classical AO system Altair. This comes together with an increase in the uniformity of the AO-corrected PSF, which improves the photometric accuracy.

2. MCAO has been demonstrated for correction in the NIR on 8-m telescopes on NGS only (MAD at the VLT) and LGS (GeMS at Gemini) and at visible wavelengths down to 430 nm for solar astronomy (Clear at the Big Bear Solar Observatory on a 1.6-m telescope). GeMS demonstrated Strehl ratio up to 40% and FWHM down to 52 mas over an FoV of $85 \times 85$ arcsec$^2$ at H band, with typical performance closer to 80 mas and 25% Strehl at K band.

3. To reach a significant fraction of the sky, MCAO relies on the use of LGSs—typically four to six to provide the multiple guide sources for tomography. GSL technology is now maturing, with turnkey lasers becoming available—e.g., TOPTICA lasers.

4. Understanding photometry and astrometry with MCAO is still very much a work in progress, but current results show depth gains of 1.5–1.7 mag for point sources over an FoV larger than 1 arcmin when compared with seeing-limited observations and astrometric stability down to 150 μas over a few hours.

5. A second generation of MCAO instruments is in the works, including NFIRAOS and MAORY, two facilities for the extremely large telescopes (respectively 35 arcsec and 20–120 arcsec FoV depending on the mode), pushing toward visible wavelengths for 8-m telescopes with MAVIS (MCAO Assisted Visible Imager and Spectrograph; 30-arcsec FoV), and the 4-m DKIST.

**FUTURE ISSUES AND DEVELOPMENTS**

1. Develop better and cheaper GSLs: (*a*) Cheaper lasers would mean that more of them could be used, reducing the tomographic error and/or enlarging the FoV; (*b*) time-gated pulsed lasers would reduce or entirely eliminate laser fratricide; and (*c*) frequency-scanned lasers would (modestly) improve the pumping efficiency of the sodium layer and make brighter LGSs.

2. Improving the efficiency and overheads of MCAO operation may seem like a mundane issue, but it is an important practical one. ESO is going a long way toward optimizing operation with the AOF (Kolb & et al. 2017), with acquisition overheads on the order of a minute.

3. Stable control of more than two DMs still has to be demonstrated in the field, even if it works on paper and in simulations.

4. Improve telemetry to optimize the MCAO control; understand and characterize performance in more detail to improve the astronomical exploitation of MCAO images. In particular, can the PSF variability in the field be improved or predicted through PSF reconstruction methods?

5. A long shot: The availability of continuous, transmissive, and achromatic 3D phase correctors would be a game changer and would allow almost limitless corrected FoVs.

## DISCLOSURE STATEMENT

The authors are not aware of any affiliations, memberships, funding, or financial holdings that might be perceived as affecting the objectivity of this review.

## ACKNOWLEDGMENTS

The authors are grateful to Marcos van Dam, Fernando Quiros-Pacheco, Dirk Schmidt, Christoph Baranec, Simone Esposito, Carlos Correia, Johan Kolb, Enrico Marchetti, Thierry Fusco, Vincent Garrel, Miska Le Louarn, and Paolo Turri for helpful comments and reading through a draft version of this paper. Thanks are extended to Markus Dirngerber for providing helpful comments on the language.